\documentclass[12pt]{article}
\usepackage{amsmath}
\usepackage{feynmf}
\def\appendix{\par\clearpage
  \setcounter{section}{0}
  \setcounter{subsection}{0}
  \@addtoreset{equation}{section}
  \def\@sectname{Appendix~}
  \def\theequation{\thesection.\arabic{equation}}
  \def\thesection{\Alph{section}}}
\makeatother
\newcommand{\be} {\begin{equation}}
\newcommand{\ee} {\end{equation}}
\newcommand{\ben} {\begin{equation*}}
\newcommand{\een} {\end{equation*}}
\renewcommand{\not}[1]{#1\!\!\!/}
\begin{document}
\numberwithin{equation}{section}
\renewcommand{\theequation}{\arabic{section}.\arabic{equation}}
\begin{titlepage}
\hskip 11cm \vbox{ \hbox{Budker INP 2003-13} \hbox{DESY  03-025} }
\vskip 0.3cm
\begin{center}
{\bf RADIATIVE CORRECTIONS TO QCD AMPLITUDES IN QUASI-MULTI-REGGE
KINEMATICS $^{~\ast}$}
\end{center}
\vskip 0.5cm
\centerline{V.S.~Fadin$^{\dag}$,
M.G.~Kozlov$^{\ddag}$ A.V.~Reznichenko$^{\dag\dag}$}
\vskip .3cm
\begin{center}
{\sl Institute for Nuclear Physics, 630090 Novosibirsk, Russia\\
and Novosibirsk State University, 630090 Novosibirsk, Russia}
\end{center}
\vskip 1cm
\begin{abstract}
Radiative corrections to  QCD amplitudes in the quasi-multi-Regge
kinematics are interesting in particular since the Reggeized form
of these amplitudes is used in the derivation of the NLO BFKL.
This form is a hypothesis which must be at least carefully
checked, if not proved. We calculate the radiative corrections in
the one-loop approximation using the $s$-channel unitarity.
Compatibility of the Reggeized form of  the amplitudes with the
$s$-channel unitarity  requires fulfillment of the set of
nonlinear equations for the Reggeon vertices. We show that these
equations are satisfied.
\end{abstract}
\vfill
\hrule \vskip.3cm
\noindent $^{\ast}${\it Work supported in part by INTAS and in
part by the Russian Fund of Basic Researches.}
\vfill $
\begin{array}{ll} ^{\dag}\mbox{{\it e-mail address:}} &
\mbox{FADIN@INP.NSK.SU}\\
^{\ddag}\mbox{{\it
e-mail address:}} &
\mbox{M.G.KOZLOV@INP.NSK.SU}\\
^{\dag\dag}\mbox{{\it e-mail address:}} &
\mbox{A.V.REZNICHENKO@INP.NSK.SU}\\
\end{array}
$
\end{titlepage}

\vfill

\eject
\section{Introduction}

In the limit of large center of mass energy {$\sqrt s$} and fixed
momentum transfer {$\sqrt {-t}$} ({Regge limit}) the most
appropriate approach for the description of  scattering amplitudes
is given by the theory of complex angular momenta ({Gribov-Regge
theory}). One of  remarkable properties of {QCD} is the
{Reggeization} of its elementary particles. Contrary to {QED},
where the electron does Reggeize in perturbation theory
\cite{GGLMZ}, but the photon remains elementary \cite{M},
 in QCD the gluon  does Reggeize
\cite{GST}-\cite{BFKL} as well as the quark \cite{FS}-\cite{KLPV}.

The phenomenon of the Reggeization is very important for  high
energy QCD. In particular, the BFKL approach \cite{BFKL} to the
description of high energy QCD processes is based on the {gluon
Reggeization}. It was assumed in this approach that  the
amplitudes with colour octets  and negative signatures in channels
with fixed (not increasing with $s$) transferred momenta have the
Reggeized form. In the leading logarithmic approximation (LLA),
when only the leading terms ($~\alpha _S\ln s)^n$ are
resummed~\cite{BFKL}, the assumption was made about  the
amplitudes in the multi-Regge kinematics (MRK). Remind that the
MRK  means large invariant masses of any pair of final state
particles and fixed transverse momenta; we include here the Regge
kinematics (RK) in the MRK  as a particular case. The Reggeized
form of these amplitudes in the LLA was proved~\cite{BLF}, so that
in this approximation the BFKL approach is completely justified.

Now the BFKL approach is developed in the next-to-leading
approximation (NLA), when the terms $~\alpha _S(\alpha _S\ln s)^n$
are also resummed. The kernel of the BFKL equation for the forward
scattering ($t=0$ and colour singlet in the $t$-channel) in the
next-to-leading order (NLO) is found ~\cite{FL98},\cite{CC98} .
The calculation of the NLO kernel for the non-forward
scattering~\cite{FF98} is not far from completion
(see~\cite{FFP99},\cite{FG}). The impact factors of
gluons~\cite{FFKPg} and quarks~\cite{FFKPq} are calculated in the
NLO and the impact factors of the physical (colour singlet)
particles are under
investigation~\cite{FM99},\cite{FIK1},\cite{BGQ},\cite{FIK2},\cite{BGK}.

The NLO results are obtained assuming  the Reggeized form both for
the amplitudes in the quasi-multi-Regge kinematics (QMRK), where a
pair of produced particles  has fixed invariant mass, and for the
MRK amplitudes in the NLA. It's clear that these assumptions  must
be at least carefully checked, if not proved. It can be done by
revision of the "bootstrap" relations ~\cite{FF98}, appearing from
the requirement of  compatibility of the Reggeized form of the
amplitudes with the $s$-channel unitarity. For the elastic
amplitudes these relations impose the bootstrap conditions on the
colour-octet impact factors and the BFKL kernel in the NLO
~\cite{FF98}. The conditions  for the impact factors of
gluons~\cite{FFKPg} and quarks~\cite{FFKPq}, as well as  for the
quark part of the kernel~\cite{FFP99}, were shown to be satisfied
at arbitrary space-time dimension $D$.  For the gluon part of the
kernel fulfillment of the bootstrap condition was proved at
$D\rightarrow 4$~\cite{FFK00}, in particular, because  this part
was available at that time only in such limit. Now it  can be
done at arbitrary $D$, since the kernel at arbitrary $D$ is
calculated~\cite{FFP00}.

Evidently, the bootstrap relations must be satisfied for all
amplitudes which were  assumed to have the Reggeized form, so that
there is an infinite set of such relations. Since the amplitudes
are expressed in terms of the gluon trajectory and a finite number
of the Reggeon vertices, it is extremely nontrivial to satisfy all
these relations. Nevertheless, it occurs that all of them can be
fulfilled if the vertices and trajectory submit to  several
bootstrap conditions~\cite{tbp}. On the other hand, the
fulfillment of all bootstrap relations secures  the Reggeized form
of the radiative corrections order by order in the perturbation
theory. On this way the proof of the Reggeization was constructed
in the LLA~\cite{BLF}. An analogous proof can be constructed in
the next-to-leading approximation (NLA) as well~\cite{tbp}.

The bootstrap relations  for the multi-particle production
amplitudes give~\cite{tbp}, in particular,  stronger restrictions
on the octet impact factors and kernel, than the relations for the
elastic amplitudes. These restrictions are known as the strong
bootstrap conditions suggested, without derivation,
in~\cite{B99,BV00}, which lead to  remarkable properties of the
colour-octet impact factors and the Reggeon
vertices~\cite{FFKP00}, that their ratio is a process-independent
function.  In the NLO this quite nontrivial property  was verified
by comparison of such ratio  for quarks and gluons~\cite{FFKP00}.
Moreover, the process-independent function mentioned above must be
the eigenfunction of the octet kernel. In the part concerning the
quark contribution to the kernel it is proved rather easily
\cite{FFP99}, \cite{BV00},\cite{BV99}. To do this for the gluon
contribution requires much more efforts, but recently it was also
done~\cite{FP02}.

In this paper we  investigate the bootstrap relations   for the
production amplitudes in the QMRK. We calculate the one-loop
radiative corrections to these amplitudes using the $s$-channel
unitarity, derive the bootstrap conditions for the production
vertices and  demonstrate  that they are fulfilled.

 The next Section contains all necessary definitions
and denotations. Then, in  Section 3, we consider the amplitudes
with a couple of particles  in the fragmentation region of one of
colliding particles. We calculate the one-loop radiative
corrections for these amplitudes and derive the bootstrap
conditions for the Reggeon vertices in the QMRK in Subsection 3.1.
In Subsections 3.2, 3.3 and 3.4 we demonstrate that these
conditions are satisfied for quark-antiquark, gluon-gluon and
quark-gluon production respectively. Next we consider production
of a couple of particles with fixed invariant mass in the central
region of rapidities. Subsection 4.1 contains the calculation of
the one-loop radiative corrections and derivation of the bootstrap
conditions. Fulfillment of these conditions is proved in
Subsections 4.2 and 4.3 for quark-antiquark and gluon-gluon
production respectively. Significance of the obtained results is
discussed in Section 5.

\section{Definitions and denotations}
Considering collisions of high energy particles $A$ and $B$ with
momenta $p_A$ and $p_B$ and masses $m_A$ and $m_B$ we introduce
light cone 4-vectors  $p_1$ and $p_2$ so that \be
 p_A=p_1+\left(m_A^2/s\right) p_2~, \;\;
 p_B=p_2+\left(m_B^2/s\right)p_1~, \;\; s=2p_1p_2\simeq (p_A+p_B)^2~,
\ee where $s$ is supposed tending to infinity, and use the Sudakov
decomposition of momenta
\begin{equation}
p=\beta  p_1+\alpha p_2 +p_{\perp }~\, , \,\,\,s\alpha \beta =p^2
-p_{\perp }^2=p^2 +\vec p^{~2}~, \label{sud}
\end{equation}
where the vector sign denotes   components of momenta transverse
to the $p_A, p_B$ plane. They are supposed to be limited (not
growing with $s$).

 According to the hypothesis of the gluon Reggeization
the amplitude of the process $A+B$\ $\rightarrow A^{\prime
}+B^{\prime }$  with a colour octet in the $t$-channel and
negative signature (that means antisymmetry under the substitution
$s\leftrightarrow u\simeq -s $ ) has the form:
\begin{equation}
{\cal A}_{AB}^{A^{\prime }B^{\prime }}=\Gamma _{A^{\prime
}A}^{c}\left[
\left( \frac{-s}{-t}\right) ^{j(t)}-\left( \frac{+s}{-t}\right) ^{j(t)}%
\right] \Gamma _{B^{\prime }B}^{c}~,
\label{Ael}
\end{equation}
where
\begin{equation}
t=q^{2}\simeq q_{\perp}^2=-\vec q^{~2}~, \;\;
q=p_{A}-p_{A^{\prime}}=p_{B^{\prime }}-p_{B}~;
\;\;\;j(t)=1+\omega(t)~;
\end{equation}
 $j(t)$ is the gluon Regge trajectory, $\Gamma _{P^{\prime
}P}^{c}$ are the vertices of the Reggeon interactions with
scattered particles,  $c$ is a colour index. The form (\ref{Ael})
represents correctly the analytical structure of the scattering
amplitude, which is quite simple in the elastic case. In  the BFKL
approach it is assumed that this form is valid in the NLA as well
as in the LLA. Remind  that in each order of perturbation theory
amplitudes with negative signature do dominate, owing to the
cancellation of the leading logarithmic terms in amplitudes with
positive signatures, which become pure imaginary in the LLA  due
to this cancellation. Note that the amplitude of the process
$A+B$\ $\rightarrow A^{\prime }+B^{\prime }$  can contain
contributions of various colour states and signatures in the
$t$-channel, so that, strictly speaking, we should indicate
somehow in the L.H.S. (\ref{Ael}) that only  the contribution of a
colour octet with  negative signature is retained. But since in
this paper we are interested only in such contributions, we have
omitted  this indication  to simplify denotations. We do the same
below considering the inelastic amplitudes, so that a colour octet
and negative signature is always assumed, without explicit
indication, in the channels with  gluon quantum numbers.

In the leading order (LO) the vertices of the Reggeon interactions
with quarks and gluons have very simple form in the helicity
basis:
\begin{equation}
 \Gamma _{P^{\prime }P}^{c} = gT^c _{P^{\prime
 }P}\delta_{\lambda_{P^{\prime}}\lambda_{P}}~,
\label{Gamma}
\end{equation}
where  $g$ is the QCD coupling constant, $T^c _{P^{\prime }P}$ are
the matrix elements of the colour group generators in
corresponding representations  and $\lambda$-s are helicities of
the partons. But we'll need  a basis-independent form of the
vertices. For quarks with momenta $p$ and $p'$ having predominant
components  along $p_1$ such form can be presented as
\begin{equation}
\Gamma^c_{Q' Q}=g \bar
u(p')t^c\frac{p\!\!\!/_2}{2pp_2}u(p)~,\label{Gamma_Q}
\end{equation}
where $t^c$ are the colour group generators in the  fundamental
representation; for  antiquarks  we  have correspondingly
\begin{equation}
\Gamma^c_{\bar Q' \bar Q}=-g \bar
v(p)t^c\frac{p\!\!\!/_2}{2pp_2}v(p')~.  \label{Gamma_bar Q}
\end{equation}
For gluons with predominant components of momenta along $p_1$
we'll use physical polarization vectors $e(p)p=e(p')p'=0$ in the
light-cone gauge $ \;\;e(p)p_2=e(p')p_2=0~,$ so that
\begin{equation}
e(p)=e(p)_{\perp}-\frac{(e(p)_{\perp}p_{\perp})}{p_2p}p_2~,\;\;
e(p')=e(p')_{\perp}-\frac{(e(p')_{\perp}p'_{\perp})}{p_2p'}p_2~,
\label{axial2}
\end{equation}
and
\begin{equation}
\Gamma^c_{G' G}=-g (e^*(p')_{\perp}e(p)_{\perp})T^c_{G'G}~,
\label{Gamma_G}
\end{equation}
with the colour generators in the adjoint representation. For
momenta with  predominant components along $p_2$ we have to \, replace 
in these formulas $p_2\rightarrow p_1$ (evidently, this
replacement in (\ref{axial2}) means change of the gauge).
The gluon trajectory in the LO is given by
\begin{equation}
\omega^{(1)}(t) =\frac{g^2N_c t}{2(2\pi )^{D-1}}
\int\frac{d^{D-2}q_1}{\vec q_1^{\:2}(\vec q-\vec q_1)^{2}}=-g^2
\frac{N_c \Gamma(1-\epsilon)}{(4 \pi)^{D/2}}
\frac{\Gamma^2(\epsilon)}{\Gamma(2\epsilon)} (\vec
q^{\:2})^\epsilon \;. \label{omega_1}
\end{equation}
Here and in the following  $N_c$ is the number of colors,
$D=4+2\epsilon$ is the space-time dimension taken different from 4
to regularize infrared divergencies; $\Gamma(x)$ is the Eueler
function.

The necessary assumption in the derivation of the BFKL equation is
the Reggeized form of the production amplitudes in the multi-Regge
kinematics (MRK), which means  large invariant masses of any pair
of final particles  and  fixed  transferred momenta. Denoting
momenta of  final particles in the process $A+B\rightarrow
P_0+P_1+...+P_{n+1}$ as $k_{i}$, $i=0\div n+1$ (see Fig.~1),
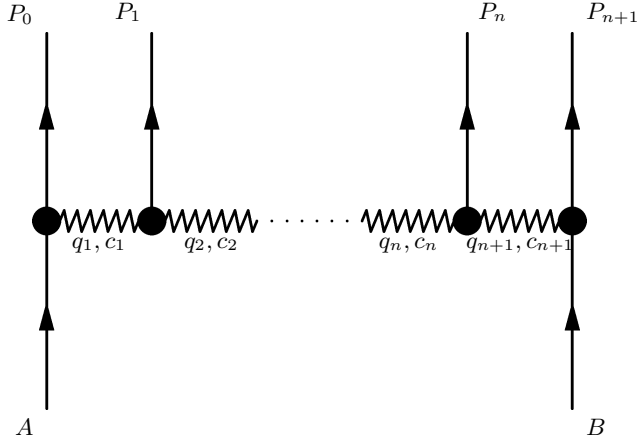
\begin{figure}
\begin{center}
\setlength{\unitlength}{1.0mm}
\begin{fmffile}{fig1}
\fontsize{9}{12pt}
\begin{fmfgraph*}(70,50)
\fmfpen{thin}
\fmfstraight
\fmftopn{r}{6}
\fmfbottomn{l}{6}
\fmf{plain_arrow}{l1,v1,r1}
\fmf{plain_arrow}{l6,v6,r6}
\fmffreeze
\fmf{zigzag,label=$q_1,, c_1$}{v1,v2}
\fmf{zigzag,label=$q_2,, c_2$}{v2,v3}
\fmf{dots}{v3,v4}
\fmf{zigzag,label=$q_n,, c_n$}{v4,v5}
\fmf{zigzag,label=$q_{n+1},, c_{n+1}$}{v5,v6}
\fmffreeze
\fmf{plain_arrow}{v2,r2}
\fmf{plain_arrow}{v5,r5}
\fmfv{label=$B$}{l6}
\fmfv{label=$A$}{l1}
\fmfv{label=$P_0$}{r1}
\fmfv{label=$P_1$}{r2}
\fmfv{label=$P_n$}{r5}
\fmfv{label=$P_{n+1}$}{r6}
\fmfv{decor.shape=circle,decor.filled=1,decor.size=0.05w}{v1,v2,v5,v6}
\end{fmfgraph*}
\end{fmffile}
\end{center}
\caption{Schematic representation of the process $A+B\rightarrow
P_0+P_1+\dots+P_{n+1}$ in the MRK. The zig-zag lines represent
Reggeized gluon exchange; the black circles denote the Regeon
vertices; $q_i$ are the Reggeon momenta, flowing from the left to
the right; $c_i$ are the colour indices.}
\end{figure}
\begin{equation}
k_{i}=\beta _{i}p_{1}+\alpha _{i}p_{2}+k_{i\perp
}~,\;\; s\alpha _{i}\beta _{i%
}=k_{i}^{2}-k_{i\perp }^{2}=k_{i}^{2}+\vec{k}_{i%
}^{~2}~, \label{sud_k_i}
\end{equation}
we can put in the MRK
\begin{equation}
\alpha _{0} \ll \alpha _{1}\dots \ll \alpha _{n}\ll \alpha
_{n+1}~, \;\;\; \beta_{n+1} \ll \beta _{n}\dots \ll \beta _{1}\ll
\beta _{0}~. \label{mrk 1}
\end{equation}
Due to Eqs. (\ref{sud_k_i}) and (\ref{mrk 1}) the squared
invariant masses
\begin{equation}
s_{i}=(k_{i-1}+k_{i})^{2}\approx s\beta_{i-1}\alpha _{i}=
\frac{\beta_{i-1}}{\beta_{i}}(k_{i}^{2}+\vec{k}_{i}^{~2})
\label{s_i}
\end{equation}
are large compared with the squared transverse momenta of produced
particles, which are of order of the squared momentum  transfers:
\begin{equation}
s_{i}\gg \vec{k}_{i}^{2}\sim \mid t_{i}\mid =\mid q_{%
i}^{2}\mid ~ , \label{mrk 2}
\end{equation}
where
\[
q_{i}=p_{A}-\sum_{j=0}^{i-1}k_{j} =
-\left(p_{B}-\sum_{j=i}^{n+1}k_{j}\right)
\approx \beta _{i}p_{1}-\alpha _{i-1}p_{2}-
\sum\limits_{j=0}^{i-1}k_{j\perp }~,
\]
\begin{equation}
t_{i}=q_{i}^{2}\approx q_{i\perp }^{2}=-\vec{q}_{i}^{~2}~,
\label{z9}
\end{equation}
and  product of  all $s_{i}$ is proportional to $s$:
\begin{equation}
\prod\limits_{i=1}^{n+1}s_{i}=s\prod\limits_{i=1}^{n}(k_{i}^{~2}+
\vec{k}_{i}^{2})~. \label{z10}
\end{equation}
The production amplitudes have a complicated analytical structure
(see, for instance,  \cite{FL93},\cite{Bart}). Fortunately, only
real parts of these amplitudes are used in the derivation of the
BFKL equation in the NLA as well as in the LLA. We restrict
ourselves also by consideration of the real parts, although it is
not explicitly indicated below.  They can be written as (see
\cite{FF98} and references therein)
\[
{\cal A}_{AB}^{\tilde{A}\tilde{B}+n} = 4(p_Ap_B)\Gamma
_{\tilde{A}A}^{c_{1}} \left[\prod_{i=1}^{n}\frac{1}{t_{i}}
\gamma_{c_{i}c_{i+1}}^{P_{i}}(q_{i},q_{i+1})\left( \frac{s_{i}}
{\sqrt{\vec{k}_{i-1}^{2}\vec{k}_{i}^{2}}}\right)^{\omega
(t_{i})}\right]
\]
\begin{equation}
\times \frac{1}{t_{n+1}}\left(
\frac{s_{n+1}}{\sqrt{\vec{k}_{n}^{2} \vec{k}_{n+1}^{2}}}\right)
^{\omega (t_{n+1})} \Gamma _{\tilde{B}B}^{c_{n+1}}~, \label{z11}
\end{equation}
where $\gamma _{c_{i}c_{i+1}}^{P_{i}}(q_{i},q_{i+1})$ are the
so-called Reggeon-Reggeon-particle (RRP) vertices, i.e. the
effective vertices for production of particles $P_i$ with momenta
$k_{i}$=$q_{i}-q_{i+1}$ in the collision of the Reggeons with
momenta $q_{i}$ and $-q_{i+1}$ and colour indices $c_{i}$ and
$c_{i+1}$. In the MRK  only gluons can be produced  with the
vertex
\begin{equation}
\gamma_{c_1c_{2}}^{G}(q_1,q_{2})
=gT_{c_1c_{2}}^{a}e^*_{\mu}(k)C^\mu (q_2,q_{1}), \label{gammaRRG}
\end{equation}
where $a$, $k=q_1-q_{2}$ and $e(k)$ are respectively   colour
index, momentum and polarization vector of the gluon,
\[
C^{\mu}(q_2,q_1) = -q^{\mu}_1-q^{\mu}_2 +
p^{\mu}_1(\frac{q_1^2}{kp_1}+2\frac{kp_2}{p_1p_2}) -
p^{\mu}_2(\frac{q_2^2}{kp_2}+2\frac{kp_1}{p_1p_2})
\]
\begin{equation}
=-q_{1\perp}^{\mu}-q_{2\perp}^{\mu}-\frac{p_1^{\mu}}{2(kp_1)}
(k_{\perp}^{2}-2q_{1\perp}^2)+\frac{p_2^{\mu}}{2(kp_2)}(k_{\perp}^{2}-2q_{2\perp}^2)
~. \label{vectorC}
\end{equation}
In the light cone gauge $e(k)p_2=0$ we have
\begin{equation}
e_\mu^*(k)C^{\mu
}(q_{2},q_1)=-2e_\perp^*(k)\left(q_{1\perp}-k_{\perp}\frac{q_{1\perp}^2}{k_{\perp}^2}\right)~.
 \label{vectorC1}
\end{equation}

In the NLA the multi-Regge form is assumed in the BFKL approach
for the production amplitudes not only in the MRK, when all
produced particles are strongly ordered in the rapidity space, but
also in the QMRK,  when a couple of two particles is produced with
rapidities of the same order. The QMRK can be obtained replacing
one of the particles $P_i$ in the MRK by this couple. Therefore
the QMRK amplitudes have the same form
(\ref{z11}) as in the MRK with one of the vertices $\gamma _{c_{i}c_{i%
+1}}^{P_{i}}$ or $\Gamma _{\tilde{P}P}^{c}$ substituted by a
vertex  for production of the couple.

If the particles  $P_1$ and $P_2$ are produced in the
fragmentation region of the particle A, we have
\begin{equation}
{\cal A}_{AB}^{\{P_1P_2\}B'}
=4(p_Ap_B)\Gamma^{c}_{\{P_1P_2\}A}\frac{s^{\omega(t)}}{t}
\Gamma^{c}_{B'B}~, \label{fragment}
\end{equation}
where now $q=p_A-k~,\;\;$ $k=k_1+k_2~, \;\;$ $k_1$ and $k_2$ are
momenta of the particles $P_1$ and $P_2$ correspondingly; for
their Sudakov parameters we have $\beta_1\sim \beta_2 \sim 1~,
\;\; \beta_1+\beta_2=1~, \;\;\alpha_1\sim \alpha_2\sim O(1/{s})~.$
The produced particles can be $gg$ or $q\bar q$ pair if the
particle $A$ is the gluon and $qg$ when the particle $A$ is the
quark.

If {rapidities} of components of the produced couple (it can be or
$gg$ or $q\bar q$ pair) are {far away}  from rapidities of
colliding particles, then it is created by two Reggeized gluons,
and its production is described by the vertices $\gamma^{Q\bar
Q}_{c_1\,c_2}(q_1, q_2)$ or $\gamma^{G_1G_2}_{c_1\,c_2}(q_1, q_2)$,
where $q_{1}, c_1$ and $-q_2, c_2$ are momenta and colour indices
of the Reggeized gluons. The amplitude ${\cal
A}_{AB}^{A'\{P_1P_2\}B'}$ describing production on the couple
$P_1$ and $P_2$ with the Sudakov parameters $\alpha_1\sim
\alpha_2\ll 1,\;\;\beta_1\sim \beta_2 \ll 1$, has the form
\begin{equation}
{\cal A}_{AB}^{A'\{P_1P_2\}B'}
=4(p_Ap_B)\Gamma^{c_1}_{A'A}\frac{s_1^{\omega(t_1)}}{t_1}
\gamma^{P_1P_2}_{c_1\,c_2}(q_1,
q_2)\frac{s_2^{\omega(t_2)}}{t_2}\Gamma^{c_2}_{B'B}~,
\label{central}
\end{equation}
where
\[
q_1=p_A-p_{A'}~,\;\; q_2=-p_B+p_{B'}~,\;\;t_i=q_i^2\simeq q_{i\perp}^2~, \;\;
\]
\begin{equation}
s_1=(p_{A'}+k)^2~,\;\;s_2=(p_{B'}+k)^2~,\;\;k=k_1+k_2~, \;\;k^2\ll
s_{1,2}\ll s~. \label{central1}
\end{equation}

Note that {because  the QMRK  in the unitarity relations  leads to
loss of the large logarithms, scales of energies in
(\ref{fragment}),(\ref{central}) are unimportant in the NLA;
moreover, the trajectory and the vertices are needed there only in
the LO}. The trajectory in this order is given by (\ref{omega_1});
the vertices are presented below. Remind that the vertices were
extracted from corresponding amplitudes in the Born approximation,
so that at the tree level Eqs.(\ref{fragment}),(\ref{central}) are
verified. What has to be checked is their energy dependence, i.e.
the Regge factors $s_i^{\omega(t_i)}$.

\section{Production in the
fragmentation region}

\subsection{One-loop radiative corrections and bootstrap conditions}

To be definite, we consider below  production in the fragmentation
region of the particle $A$. In this section we use denotations
$s_1=(p_{B'}+p_{P_{1}})^2$ and $s_2=(p_{B'}+p_{P_{2}})^2$. Note
that here $s_1\sim s_2 \sim s$, contrary to the case of production
in the central region of rapidities. In the radiative corrections
to the amplitude ${\cal A}_{AB}^{\{P_1P_2\}B'}$ we have to retain
only large logarithmic terms, not making difference between $\ln
s$, $\ln s_1$ and $\ln s_2$.
  Therefore the corrections  can be calculated using the $s$-channel
 unitarity in the same way as it was done for the elastic scattering
amplitudes in the LLA \cite{BFKL}. The large logarithms are
defined by  the discontinuities of the amplitude ${\cal
A}_{AB}^{\{P_1P_2\}B'}$ in the channels  $s$, $s_1$ and $s_2$, and
we find them using the unitarity relations in these channels.

Let start with the $s$-channel discontinuity. In the one-loop
approximation the intermediate states in the unitarity relation
can be only two-particle states, so that we have (see Fig.~2~a)
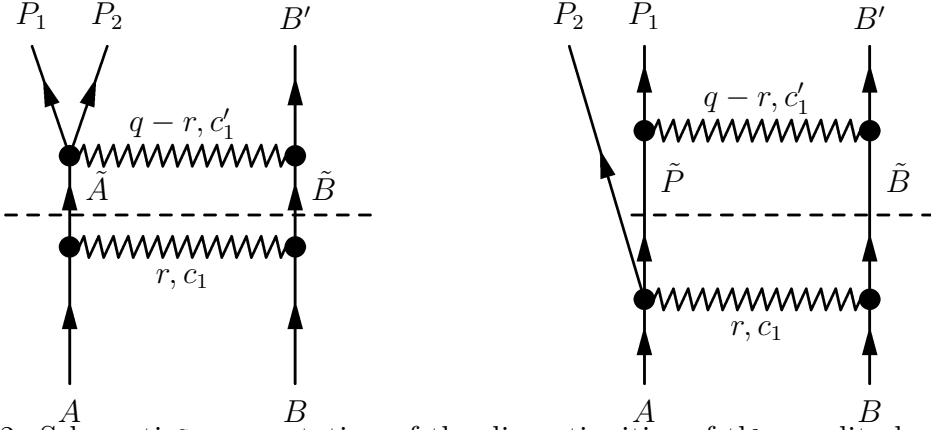
\begin{figure}
\begin{minipage}[t]{75mm}
\begin{center}
\setlength{\unitlength}{1.0mm}
\begin{fmffile}{fig2a}
\begin{fmfgraph*}(50,45)
\fmfpen{thin}
\fmfstraight
\fmftopn{t}{11}
\fmfbottomn{b}{6}
\fmfright{r}
\fmfleft{l}
\fmf{dashes}{r,l}
\fmf{plain_arrow,tension=1.6}{b2,v1}
\fmf{plain_arrow,label.angle=-90,tension=2.4}{v1,v2}
\fmf{plain_arrow}{v2,t2}
\fmf{plain_arrow}{v2,t4}
\fmf{plain_arrow,tension=0.8}{b5,v3}
\fmf{plain_arrow,tension=1.2}{v3,v4}
\fmf{plain_arrow}{v4,t9}
\fmffreeze
\fmf{zigzag,label=$r,, c_1$,label.dist=3mm}{v1,v3}
\fmf{zigzag,label=$q-r,, c'_1$}{v2,v4}
\fmfv{label=$A$,label.angle=-90}{b2}
\fmfv{label=$B$,label.angle=-90}{b5}
\fmfv{label=$B'$,label.angle=90}{t9}
\fmfv{label=$P_1$,label.angle=90}{t2}
\fmfv{label=$P_2$,label.angle=90}{t4}
\fmfv{decor.shape=circle,decor.filled=1,decor.size=0.05w}{v1,v2,v3,v4}
\fmf{phantom,label=$a$,label.side=right,label.dist=6mm }{b3,b4}
\fmfv{label=$\tilde{A}$,label.angle=-45,label.dist=3mm}{v2}
\fmfv{label=$\tilde{B}$,label.angle=-45,label.dist=3mm}{v4}
\end{fmfgraph*}
\end{fmffile}
\end{center}
\end{minipage}
\begin{minipage}[t]{75mm}
\begin{center}
\setlength{\unitlength}{1.0mm}
\begin{fmffile}{fig2b}
\begin{fmfgraph*}(50,45)
\fmfpen{thin}
\fmfstraight
\fmfbottomn{b}{6}
\fmftopn{t}{6}
\fmfright{r}
\fmfleft{l}
\fmf{plain_arrow}{b2,v1,vm1}
\fmf{plain,label=$\tilde{P}$}{vm1,v2}
\fmf{plain_arrow}{v2,t2}
\fmf{plain_arrow}{b5,v3,vm2}
\fmf{plain,label=$\tilde{B}$}{vm2,v4}
\fmf{plain_arrow}{v4,t5}
\fmffreeze
\fmf{zigzag,label=$r,, c_1$,label.dist=3mm}{v1,v3}
\fmf{zigzag,label=$q-r,, c'_1$,label.side=left}{v2,v4}
\fmffreeze                
\fmf{plain_arrow}{v1,t1}
\fmf{phantom}{l,vd1}
\fmf{dashes,tension=0.2}{vd1,r}
\fmfv{label=$P_2$,label.angle=90}{t1}
\fmfv{label=$P_1$,label.angle=90}{t2}
\fmfv{label=$B'$,label.angle=90}{t5}
\fmfv{label=$A$,label.angle=-90}{b2}
\fmfv{label=$B$,label.angle=-90}{b5}
\fmfv{decor.shape=circle,decor.filled=1,decor.size=0.05w}{v1,v2,v3,v4}
\fmf{phantom,label=$b$,label.side=right,label.dist=6mm}{b3,b4}
\end{fmfgraph*}
\end{fmffile}
\end{center}
\end{minipage}
\caption{Schematic representation of the discontinuities of the
amplitude  ${\cal
 A}_{AB}^{\{P_1P_2\}B'}$: a -in the $s$-channel; b - in the
 $s_1$-channel.}
\end{figure}

\begin{equation}
\Im_s \,{\cal A}_{AB}^{\{P_1P_2\}B'}=\frac 1{2}\sum_{\{\tilde A
\tilde B\}}\int {\cal A}^{\tilde A \tilde B}_{AB} {\cal
A}^{\{P_1P_2\}B'}_{\tilde A \tilde B} d\Phi _{\tilde A \tilde B},
\label{Imsfragment}
\end{equation}
where the sum $\sum_{\{\tilde A \tilde B\}}$ is over  over all
discrete quantum numbers of the particles $\tilde A$ and $\tilde
B$, $d\Phi_{\tilde A \tilde B }$ is  their phase space element.
Here and in the following we use the Hermitian property of the
Born amplitudes \be \left({\cal A}^{f}_{i}\right)^*={\cal
A}^{i}_{f}~. \ee In the  region which gives  a leading (growing as
$s$) contribution  to the imaginary part
\begin{equation}
d\Phi _{\tilde A \tilde B} =(2\pi )^D \delta^{(D)}
(p_A+p_B-p_{\tilde A}-p_{\tilde B})\frac{d^{D-1}p_{\tilde
A}}{2\epsilon_{\tilde A}(2\pi )^{D-1}}\frac{d^{D-1}p_{\tilde
B}}{2\epsilon_{\tilde B}(2\pi)^{D-1}}
=\frac{d^{D-2}r_{\perp}}{2s(2\pi )^{D-2}}~.  \label{rho2}
\end{equation}
Here and below $r_\perp$ is the transverse part of the momentum
transfer $p_{\tilde B}-p_{B}$.  Note that for production in the
fragmentation region the Sudakov parameters $\alpha$ and $\beta$
for the momentum transfer $p_{\tilde B}-p_{B}$ are $\sim 1/s$, so
that $p_{\tilde B}-p_{B} \simeq r_\perp$. For production in the
central region it is not always correct.

The imaginary parts in the $s_{1,2}$-channels are calculated quite
analogously. Take the $s_1$-channel. Denoting intermediate
particles in the unitarity relation in this channel $\tilde P$ and
$\tilde B$, we obtain (see Fig.~2~b)
\begin{equation}
\Im_{(p_{B'}+p_{P_{1}})^2} {\cal A}_{AB}^{\{P_1P_2\}B'}= \frac
1{2}\sum_{\{\tilde P \tilde B\}}\int {\cal A}^{\{\tilde P
P_2\}\tilde B}_{AB} {\cal A}^{P_1 B'}_{ \tilde P \tilde B} d\Phi
_{ \tilde P \tilde B}, \label{Ims1fragment}
\end{equation}
with
\begin{equation}
d\Phi _{\tilde P \tilde B} =(2\pi )^D \delta^{(D)}
(p_{P_1}+p_{B'}-p_{\tilde P}-p_{\tilde B})\frac{d^{D-1}p_{ \tilde
P}}{2\epsilon_{\tilde P}(2\pi )^{D-1}}\frac{d^{D-1}p_{\tilde
B}}{2\epsilon_{\tilde B}(2\pi)^{D-1}}
=\frac{d^{D-2}r_{\perp}}{2(p_{B'}+p_{P_{1}})^2(2\pi )^{D-2}}~.
\label{rho2i}
\end{equation}
The $s_2$-channel imaginary part is obtained from
(\ref{Ims1fragment}), (\ref{rho2i}) by the substitution
$P_1\leftrightarrow P_2$.

Since we don't make difference between $\ln s$, $\ln s_1$ and $\ln
s_2$, we need only sum of the imaginary parts in the $s$, $s_1$
and $s_2$ channels. Using (\ref{Ael}) and (\ref{fragment}) in the
Born approximation for the amplitudes in (\ref{Imsfragment}),
(\ref{Ims1fragment}) we obtain for the sum
\begin{equation}
\Im {\cal A}_{AB}^{\{P_1P_2\}B'}=\frac{s}{(2\pi
)^{D-2}}\int\frac{d^{D-2}r_{\perp}}{r^2_{\perp}(q-r)^2_\perp}
\sum_{\{i\}}\Gamma^{c_1}_{\{i\}
A}\Gamma^{c_1^\prime}_{\{P_1P_2\}\{i\}} \sum_{\{\tilde
B\}}\Gamma^{c_1}_{\tilde B B}\Gamma^{c_1^\prime }_{B'\tilde B}~,
\label{}
\end{equation}
where the sum over $\{i\}$  is performed {over all possible
intermediate states and their quantum numbers}. If $\{i\}$
 contains two particles, one of them must be $P_1$ or $P_2$;
 in this case corresponding subscript in
 $\Gamma^{c_1^\prime}_{\{P_1P_2\}\{i\}}$ can be omitted.

Remind that we assume everywhere  projection on   a colour octet
and  negative signature in the $t$-channel. Performing this
projection explicitly by  the projection  operator $\hat{\cal
{P}}_{8_a}$,
\begin{equation}
\langle c_{1}c_{1}^{\prime }|\hat{\cal
{P}}_{8_a}|c_{2}c_{2}^{\prime }\rangle
=\frac{f_{c_{1}c_{1}^{\prime }c}f_{c_{2}c_{2}^{\prime }c}}{N_c}\ ,
\label{project}
\end{equation}
where $f_{abc}$ are the structure constants of the colour group,
and using the bootstrap property of  the LO vertices
\begin{equation}
f_{c_{1}c_{1}^{\prime }c}\sum_{\{\tilde B\}} \Gamma^{c_1}_{\tilde
B B}\Gamma^{c_1^\prime }_{B'\tilde B}=-ig\frac{N_c}{2}
\Gamma^{c}_{B' B}~,  \label{boot lo}
\end{equation}
which is easily derived from (\ref{Gamma}), we get
\begin{equation}
\Im {\cal A}_{AB}^{\{P_1P_2\}B'}=\frac{s}{t}\left(-\pi
\frac{gt}{(2\pi)^{D-1}}
\int\frac{d^{D-2}r_{\perp}}{r^2_{\perp}(q-r)^2_{\perp}}if_{c_{1}c_{1}^{\prime}c}
\sum_{\{i\}}\Gamma^{c_1}_{\{i\}
A}(r_\perp)\Gamma^{c_1^\prime}_{\{P_1P_2\}\{i\}}(q_\perp-r_\perp)\right)
\Gamma^{c}_{B' B}~. \label{im}
\end{equation}
Here we indicate  explicitly dependence of the Reggeon vertices on
momentum transfer. Remind that the sum over $\{i\}$ is performed
{over all possible intermediate states and their quantum numbers}.
All vertices here are taken in the leading order, so that if an
intermediate state contains two particles, one of them must be the
same as in the final state; another changes its transverse
momentum and colour state, but its helicity is conserved. The real
part of the one-loop contribution to the amplitude can be restored
from the imaginary part by the substitution (cf.(\ref{Ael}))
\begin{equation}
-\pi \;\rightarrow \;2\ln s~.  \label{sub s}
\end{equation}
Therefore,  comparing (\ref{im}) with the first order term in the
expansion of (\ref{fragment}) with account of (\ref{omega_1}), we
see that the one-loop correction calculated above is compatible
with the Reggeized form (\ref{fragment}) only if
\begin{equation}
\int \frac{d^{D-2}r_{\perp}}{r_{\perp} ^2(q-r)^{2}_\perp}
\frac{if^{cc_1c_1^{\prime}}}{N_c}\sum_{\{i\}}
\Gamma^{c_1}_{{\{i\}}A}(r_\perp)
\Gamma^{c_1^{\prime}}_{\{P_1P_2\}\{i\}}(q_\perp-r_\perp)
=\frac{g}{2}\Gamma^c_{\{P_1P_2\} A}(q_\perp) \int
\frac{d^{D-2}r_{\perp}}{r_{\perp} ^2(q-r)^{2}_\perp}~. \label{boot
qmrk}
\end{equation}

Eq. (\ref{boot qmrk}) gives  the bootstrap conditions for the
Reggeon vertices of  two-particle production  in the fragmentation
region. In the next subsections we show that they are satisfied.

\subsection{Quark-antiquark production}

To produce a $q\bar q$ pair  the particle $A$ must be a gluon. Let
$p_A=p_1$, $a$ is the colour index of the initial gluon, $k_{1}$
and $k_2$ are the quark and antiquark momenta respectively,
\begin{equation}
k_{1, 2} = \beta_{1, 2}\:p_1 + \frac{m^2 + \vec k_{1,
2}^{\:2}}{s\beta_{1, 2}} p_2 + k_{1, 2\perp}~,\;\;\;
k_{1\perp}+k_{2\perp}+q_{\perp}=0~,
 \label {}
\end{equation}
$m$ is the quark mass. The intermediate states ${\{i\}}$ in (\ref{boot qmrk}) can be:\\
1) {one-gluon state} with  momentum
$p_{\tilde A}=p_1-r$;\\
2) {$q\bar q$ state} with  quark and antiquark momenta
 $k_1^{\prime}=k_1+q-r$ and $k_2$ respectively;\\
 3) {$q\bar q$
state} with quark and antiquark momenta respectively $k_1$ and
$k_2^{\prime}=k_2+q-r$.\\
Apart from the "elastic" vertices
(\ref{Gamma_G}),(\ref{Gamma_Q}),(\ref{Gamma_bar Q}) the bootstrap
condition contains only the Reggeon vertex for $q\bar q$
production, which can be found in \cite{FFKPg}. In general case,
when the pair is produced by the gluon $G$ with momentum $k=\beta
p_1 + \vec k^{\:2}/({\beta s})p_2 + k_{\perp}~$, the vertex can be
presented as
\[
\Gamma^c_{\{Q\bar Q\}G} =  \left( t^at^c \right)_{i_1i_2}
\left({\cal A}((k_1-x_1k)_\perp)-{\cal
A}((x_2k_1-x_1k_2)_\perp)\right)
\]
\begin{equation}
-\left(t^ct^a \right)_{i_1i_2}\left({\cal A}((-k_2+x_2k)_\perp)
-{\cal A} ((x_2k_1-x_1k_2)_\perp) \right)~, \label{Gamma gqq}
\end{equation}
where $x_{1,2}=\beta_{1,2}/\beta, \;\;x_1+x_2=1$, $\;\;i_1, i_2$
are quark and antiquark colour indices,   $a$ is the colour index
of the gluon $G$. The amplitudes ${\cal A}(p_\perp)$ in the
light-cone gauge (\ref{axial2}) are  rather simple: \be {\cal
A}(p_\perp) =\frac{g^2}{p_\perp^2-m^2} \bar
u(k_1)\frac{p\!\!\!/_B}{\beta s}\biggl(x_1 e\!\!\!/_{\perp}
p\!\!\!/_{\perp} - x_2p\!\!\!/_{\perp}  e\!\!\!/_{\perp} -
e\!\!\!/_{\perp}m \biggr)v(k_2)~. \ee Here $e$ is the gluon
polarization vector, $u(k_1)$ and $v(k_2)$ are the spin wave
functions of the quark and antiquark  respectively.

With the vertices (\ref{Gamma_G}),(\ref{Gamma_Q}),(\ref{Gamma_bar
Q}) and (\ref{Gamma gqq}) the contribution of either of the three
intermediate states to the integrand in  L.H.S. of (\ref{boot
qmrk})
is readily calculated and we obtain correspondingly\\
{1)}
\[
\frac{igf^{cc_1c_1^{\prime}}}{N_c}T^{c_1}_{a^\prime a}\left[
\left( t^{a^\prime}t^{c_1^{\prime}}\right)_{i_1i_2} \left({\cal
A}((k_1+x_1r)_\perp)-{\cal A}((x_2k_1-x_1k_2)_\perp)\right)\right.
\]
\begin{equation}
\left.-\left(t^{c_1^{\prime}}t^{a^\prime}
\right)_{i_1i_2}\left({\cal A}((-k_2-x_2r)_\perp) -{\cal
A}((x_2k_1-x_1k_2)_\perp) \right)\right]~,\label{one-gluon}
\end{equation}
{2)}
\[
\frac{igf^{cc_1c_1^{\prime}}}{N_c}\left[\left(t^{c_1^{\prime}}
t^{a}t^{c_1} \right)_{i_1i_2} \left({\cal A}((-k_2-r)_\perp)-{\cal
A}((-k_2-x_2r)_\perp)\right)\right.
\]
\begin{equation}
\left.-\left(t^{c_1^{\prime}} t^{c_1}t^{a}
\right)_{i_1i_2}\left({\cal A}((-k_2)_\perp) -{\cal
A}((-k_2-x_2r)_\perp) \right)\right]~,\label{two-gluon 1}
\end{equation}
and\\
{3)}
\[
- \frac{igf^{cc_1c_1^{\prime}}}{N_c}\left[\left(
t^{a}t^{c_1}t^{c_1^{\prime}} \right)_{i_1i_2} \left({\cal
A}((k_1)_\perp)-{\cal A}((k_1+x_1r)_\perp)\right)\right.
\]
\begin{equation}
\left.-\left(t^{c_1}t^{a}t^{c_1^{\prime}}
\right)_{i_1i_2}\left({\cal A}((k_1+r)_\perp) -{\cal
A}((k_1+x_1r)_\perp) \right)\right]~.\label{two-gluon 2}
\end{equation}
It's not difficult to see from these expressions  that  the terms
with ${\cal A}((k_1+x_1r)_\perp)$ are cancelled  before
integration, due to the commutation relations between $t^i$, as
well as the terms with ${\cal A}((-k_2-x_2r)_\perp)$. As for the
terms with ${\cal A}((k_1+r)_\perp)$ and ${\cal
A}((-k_2-r)_\perp)$, they cancel each other as a result of
integration, due to invariance of the integration measure
$d^{D-2}r_{\perp}/\left({r_{\perp} ^2(q-r)^{2}_\perp}\right)$ with
respect to the substitution $(k_1+r)_\perp\leftrightarrow
(-k_2-r)_\perp$, with account of
$k_{1\perp}+k_{2\perp}+q_{\perp}$=0. A simple colour algebra shows
that the remaining terms gather into $(g/2)\Gamma^c_{\{Q\bar Q\}
A}$, where $A$ is a gluon with momentum $p_A=p_1$ (see (\ref{Gamma
gqq})), that makes evident that the bootstrap condition (\ref{boot
qmrk}) is satisfied.

\subsection{Two-gluon production}

The case of two-gluon production can be considered quite
similarly. Again the particle $A$ must be a gluon. Using the same
denotations as before, with the difference that  $k_{1}$ and $k_2$
now are the momenta of the produced gluons (so that $m$ is
replaced by $0$), $i_{1}$ and $i_2$ are their colour indices.
Denoting their polarization vectors in the light-cone gauge
(\ref{axial2}) $e_{1}$ and $e_2$, we can present  the vertex
$\Gamma^c_{\{G_1G_2\}G}$ of two-gluon production \cite{FFKPg} in
the same form as (\ref{Gamma gqq})
\[
\Gamma^c_{\{G_1G_2\} G} =  \left( T^aT^c \right)_{i_1i_2}
\left({\cal A}((k_1-x_1k)_\perp)-{\cal
A}((x_2k_1-x_1k_2)_\perp)\right)
\]
\be
 -\left(T^cT^a \right)_{i_1i_2}\left({\cal
A}((-k_2+x_2k)_\perp) -{\cal A} ((x_2k_1-x_1k_2)_\perp)
\right)~,\label{Gamma ggg } \ee
 where  the amplitudes ${\cal A}(p_\perp)$ now have
the form: \be {\cal A}(p_\perp) =\frac{2g^2}{p_\perp^2} \biggl[
x_1x_2 \left( e_{1\perp} ^*e_{2\perp}^* \right)\left(
e_{\perp}{p}_{\perp} \right) - x_1\left( e_{1\perp}^*e_{\perp}
\right)\left( e_{2\perp}^*{p}_{\perp} \right) - x_2\left(
e_{2\perp}^*e_{\perp} \right)\left( e_{1\perp}^*{p}_{\perp}
\right) \biggr]~. \ee
The intermediate states are now:\\
1) one-gluon state with gluon momentum
$p_{\tilde A}=p_1-r$;\\
2) {two-gluon  state} with gluon momenta $k_1^{\prime}=k_1+q-r$
and $k_2$;\\
3) {two-gluon  state} with gluon momenta $k_1$ and
$k_2^{\prime}=k_2+q-r$.

It is easy to see  that the contributions of these  states
 to  the integrand  in  L.H.S. of (\ref{boot qmrk})  are given by the
 {same formulas} (\ref{one-gluon})-(\ref{two-gluon 2})  as for
the case of  quark-antiquark production,  with the only difference
that the colour group generators are taken  {not in the
fundamental, but in the adjoint representation}. Since in {the
proof} of fulfillment of the bootstrap conditions only the
commutation relations of the generators were used, the proof {can
be applied} to the case of two-gluon production as well as to
$q\bar q$ production.

\subsection{Quark-gluon production}

In the case of {quark-gluon production} (when the particle $A$ is a
quark)  the bootstrap condition  can be considered in the same
way. Let now $k$ is the momentum of  incoming quark, $k_1$ and
$k_2$ are  the momenta of  final quark and gluon correspondingly.
Note that $k^2=k_1^2=m^2$, so that
\[
k=\beta p_1 + \frac{\vec k^{\:2}+m^2}{\beta s}p_2 +
k_{\perp}~,\;\;
\]
\begin{equation}
k_1=\beta_1 p_1 + \frac{\vec k_1^{\:2}+m^2}{\beta_1 s}p_2 +
k_{1\perp}~,\;\;k_2=\beta_2 p_1 + \frac{\vec k_2^{\:2}}{\beta_2
s}p_2 + k_{2\perp}~.
\end{equation}
Then from \cite{FFKPq}
one can obtain
\[
\Gamma^c_{\{QG\} Q} =  \left( t^at^c \right)_{i_1i_2} \left({\cal
A}((x_2k_1-x_1k_2)_\perp)-{\cal A}((k_1-x_1k)_\perp)\right)
\]
\be -\left(t^ct^a \right)_{i_1i_2}\left({\cal
A}((-k_2+x_2k)_\perp) -{\cal A} ((k_1-x_1k)_\perp)
\right)~,\label{Gamma QGQ} \ee where $i_1$ and $i_2$ are now the
colour indices of the outgoing and incoming quarks,  $a$ is the
colour index of the produced gluon $G$, and the amplitudes ${\cal
A}$ now have the form:
\begin{equation}
{\cal A}(p_\perp) =-\frac{g^2}{p_\perp^2-x_2^2m^2} \bar
u(k_1)\frac{p\!\!\!/_B}{\beta s}\biggl(x_1 {e\!\!\!/}^*_{\perp}
p\!\!\!/_{\perp} +p\!\!\!/_{\perp}  {e\!\!\!/}^*_{\perp} +
{e\!\!\!/}^*_{\perp}x_2^2m \biggr)u(p)~.
\end{equation}
Possible intermediate states are now: \\
1) {one-quark state} with quark momentum $p_{\tilde A}=p_A-r$; its
contribution to the integrand in the L.H.S. of the bootstrap
equation is
\[
\frac{igf^{cc_1c_1^{\prime}}}{N_c}\left[ \left(
t^{a}t^{c_1^{\prime}}t^{c_1} \right)_{i_1i_2} \left({\cal
A}((x_2k_1-x_1k_2)_\perp)-{\cal A}((k_1+x_1r)_\perp)\right)
\right.
\]
\be \left.-\left(t^{c_1^{\prime}}t^{a}t^{c_1} \right)_{i_1i_2}
\left({\cal A}((-k_2-x_2r)_\perp) -{\cal A}((k_1+x_1r)_\perp)
\right)\right]~; \ee 2) {quark-gluon state} with  quark and gluon
momenta $k_1^{\prime}=k_1+q-r$ and $k_2$ correspondingly; it gives
\[
\frac{igf^{cc_1c_1^{\prime}}}{N_c}\left[\left(t^{c_1^{\prime}}
t^{a}t^{c_1} \right)_{i_1i_2} \left({\cal
A}((-k_2-x_2r)_\perp)-{\cal A}((-k_2-r)_\perp)\right)\right.
\]
\be \left.-\left(t^{c_1^{\prime}} t^{c_1}t^{a}
\right)_{i_1i_2}\left({\cal A}((-k_2)_\perp) -{\cal
A}((-k_2-r)_\perp) \right)\right]~, \ee
and \\
3) {quark-gluon state} with  quark and gluon  momenta $k_1$ and
$k_2^{\prime}=k_2+q-r$; it contributes
\[
\frac{igf^{cc_1c_1^{\prime}}}{N_c} T^{c_1^\prime}_{aa^\prime }
\left[ \left( t^{a^\prime}t^{c_1} \right)_{i_1i_2} \left({\cal
A}((k_1+x_1r)_\perp)-{\cal A}((k_1)_\perp)\right)\right.
\]
\be \left.-\left(t^{c_1}t^{a^{\prime}} \right)_{i_1i_2}\left({\cal
A}((k_1+r)_\perp) -{\cal A}((k_1)_\perp) \right)\right]~. \ee

As well as in the case of $q\bar q$ production, it's not difficult
to see  that the terms with ${\cal A}((k_1+x_1r)_\perp)$ and
${\cal A}((-k_2-x_2r)_\perp)$ are cancelled before integration,
due to  colour algebra;  the terms with ${\cal A}((k_1+r)_\perp)$
and ${\cal A}((-k_2-r)_\perp)$ cancel each other as a result of
integration,  and  the remaining terms give $(g/2)\Gamma^c_{\{GQ\}
 Q}$, where $Q$ is a quark with momentum $p_A=p_1+(m^2/s) p_2$ (see (\ref{Gamma QGQ})).

It completes the proof that  the bootstrap conditions (\ref{boot
qmrk}) are satisfied.

We have considered here the case of $qg$ production.  QCD
invariance under the charge conjugation secures that the bootstrap
condition is fulfilled also for  $\bar q g$ production.

\section{Production in the
central region}

\subsection{One-loop radiative corrections and bootstrap conditions}

Seing that only large logarithmic terms in the radiative
corrections to the amplitude ${\cal A}_{AB}^{A'\{P_1P_2\}B'}$ must
be retained, the corrections again can be calculated using the
$s$-channel unitarity, as it was done for  gluon production in the
MRK in the LLA \cite{BFKL}. The logarithmic terms in the real part
of the amplitude  are obtained from  the imaginary parts,
connected with the discontinuities of the amplitude in channels
with  great (tending to infinity when $s\rightarrow \infty$)
invariants, by the substitution (\ref{sub s}), with corresponding
invariant instead of $s$. Production of two particles with fixed
invariant mass instead of one leads only to technical
complications connected with existence of larger number of such
invariants, analogously to the case of two particles in the
fragmentation region compared with  elastic scattering.

Let momenta of the produced particles $P_1$ and $P_2$  be $k_1$
and  $k_2$ with $k_1+k_2=k=q_1-q_2\;$; $q_1=p_A-p_{A'}$ and
$q_2=p_{B'}-p_{B}$ are transferred momenta; note that we can
neglect by a component of $q_1\; (q_2)$ along $p_2\; (p_1)$, so
that
\begin{equation}
q_1=\beta p_1 +q_{1\perp}~, \;\; q_2=-\alpha p_2 +q_{2\perp}~,
\;\; s\alpha\beta=\vec k^{~2}~. \end{equation} In the case
of production of one particle with momentum $k$ in the MRK the
large logarithms were defined  by the discontinuities in the
channels $s_1=(p_{A'}+k)^2$, $s_2=(p_{B'}+k)^2$, $s$ and
$(p_{A'}+p_{B'})^2$.  Now we have more invariants which are great;
but they can be divided into three groups of invariants of the
same order ($\sim s_1$, $\sim s_2$ and $\sim s$). Evidently, we
have to calculate discontinuities in channels of all these
invariants. Since we don't differ logarithms of invariants of one
order, the real parts of the amplitude related to discontinuities
in channels of invariants $\sim s_a$ ($ s_a$ can be $s_1,\, s_2$
or $s$) are obtained from the imaginary parts by the substitution
(\ref{sub s}) with $s\rightarrow s_a$. Note that with our accuracy
$\ln s =\ln s_1+\ln s_2$, therefore only two large logarithms in
the real part can be considered as independent. We choose as
independent $\ln s_1$ and $\ln s_2$. To calculate the contribution
with $\ln s_1$ ($\ln s_2$) in the real part we have to find the
sum of the imaginary parts in the channels with invariants of
order $s_1$ ($ s_2$) and of order of $s$ and then to make the
substitution (\ref{sub s}) with $s_1$ ($ s_2$) instead of $s$.

Therefore,  to find  the terms with $\ln s_2$ in the real part we
need to calculate  the imaginary parts in the channels
$s_2=(p_{B'}+k)^2$, $ s_{21}=(p_{B'}+k_1)^2$, $s_{22}=(p_{B'}+k_2)^2$,
$s=(p_{A}+p_{B})^2$, $s'=(p_{A'}+p_{B'})^2$, $s'_1=(p_{A'}+k_1+p_{B'})^2$ and
$s'_2=(p_{A'}+k_2+p_{B'})^2$, schematically shown in Figs. 3 a-g. Let
us represent the sum of the imaginary parts as
\begin{equation}
\Im {\cal A}_{AB}^{A'\{P_1P_2\}B'}=s\Gamma^{c_1}_{A'
A}\frac{1}{t_1}\left(-\pi \frac{gt_2}{(2\pi
)^{D-1}}\int\frac{d^{D-2}r_{\perp}}{r^2_{\perp}(q_2-r)^2_\perp}{\cal
F}^{P_1P_2}_{c_1c_2}(q_1,q_2,r_\perp)\right)\frac{1}{t_2}
\Gamma^{c_2}_{B' B}~. \label{Im central}
\end{equation}
Below a possibility of such representation (which could be clear
for an advanced reader) is shown  and the contributions to ${\cal
F}^{P_1P_2}_{c_1c_2}(q_1,q_2,r_\perp)$ from the imaginary parts in
each of the channels are found.
\begin{figure}
\begin{center}
\begin{minipage}[t]{54mm}
\setlength{\unitlength}{1.0mm}
\begin{fmffile}{fig3a}
\fontsize{8}{12pt}
\begin{fmfgraph*}(50,45)
\fmfpen{thin}
\fmfstraight
\fmfbottomn{b}{6}
\fmftopn{t}{6}
\fmfright{r}
\fmfleft{l}
\fmf{dashes}{r,l}
\fmf{phantom}{b1,t1}
\fmf{phantom}{b6,t6}
\fmf{plain,tension=2}{b2,v1}
\fmf{plain,tension=1}{v1,v2}
\fmf{plain_arrow}{v2,t2}
\fmf{plain,tension=2}{b5,v3}
\fmf{plain,tension=1}{v3,v4}
\fmf{plain_arrow}{v4,t5}
\fmffreeze
\fmf{zigzag,label=$q_1,, c_1 $}{v1,v5}
\fmf{zigzag,label=$r,, i$}{v5,v3}
\fmffreeze
\fmf{curly,tension=2,label.side=left,label=$k'$}{v5,v6}
\fmf{plain_arrow}{t4,v6,t3}
\fmffreeze
\fmf{zigzag,label=$q_2-r,,j$,label.side=left}{v6,v4}
\fmfv{label=$k_1$}{t3}
\fmfv{label=$-k_2$}{t4}
\fmfdot{v1,v3,v4}
\fmfv{decor.shape=circle,decor.filled=1,decor.size=0.05w}{v5,v6}
\fmfv{label=$A'$,label.angle=90}{t2}
\fmfv{label=$B'$,label.angle=90}{t5}
\fmfv{label=$A$,label.angle=-90}{b2}
\fmfv{label=$B$,label.angle=-90}{b5}
\fmf{phantom,label=$ \textmd{\large a} $,label.side=right,label.dist=6mm}{b3,b4}
\end{fmfgraph*}
\end{fmffile}
\end{minipage}
\begin{minipage}[t]{54mm}
\setlength{\unitlength}{1.0mm}
\begin{fmffile}{fig3b}
\fontsize{8}{12pt}
\begin{fmfgraph*}(50,45)
\fmfpen{thin}
\fmfstraight
\fmfbottomn{b}{6}
\fmftopn{t}{6}
\fmfright{r}
\fmfleft{l}
\fmf{dashes}{r,l}
\fmf{phantom}{b1,t1}
\fmf{phantom}{b6,t6}
\fmf{plain,tension=2}{b2,v1}
\fmf{plain,tension=1}{v1,v2}
\fmf{plain_arrow}{v2,t2}
\fmf{plain,tension=2}{b5,v3}
\fmf{plain,tension=1}{v3,v4}
\fmf{plain_arrow}{v4,t5}
\fmffreeze
\fmf{zigzag,label=$q_1,, c_1$}{v1,v5}
\fmf{zigzag,label=$r,, i$}{v5,v3}
\fmffreeze
\fmf{plain_arrow}{t3,v5}
\fmf{plain_arrow,label=$k'_1 \qquad \qquad$,label.side=right}{v5,v6}
\fmf{plain_arrow}{v6,t4}
\fmffreeze
\fmf{zigzag,label=$q_2-r,,j$,label.side=left}{v6,v4}
\fmfdot{v1,v3,v4}
\fmfv{label=$-k_2$,label.angle=90}{t3}
\fmfv{label=$k_1$,label.angle=90}{t4}
\fmfv{decor.shape=circle,decor.filled=1,decor.size=0.05w}{v5,v6}
\fmfv{label=$A'$,label.angle=90}{t2}
\fmfv{label=$B'$,label.angle=90}{t5}
\fmfv{label=$A$,label.angle=-90}{b2}
\fmfv{label=$B$,label.angle=-90}{b5}
\fmf{phantom,label=$\textmd{\large b} $,label.side=right,label.dist=6mm}{b3,b4}
\end{fmfgraph*}
\end{fmffile}
\end{minipage}
\begin{minipage}[t]{54mm}
\setlength{\unitlength}{1.0mm}
\begin{fmffile}{fig3c}
\fontsize{8}{12pt}
\begin{fmfgraph*}(50,45)
\fmfpen{thin}
\fmfstraight
\fmfbottomn{b}{6}
\fmftopn{t}{6}
\fmfright{r}
\fmfleft{l}
\fmf{dashes}{r,l}
\fmf{phantom}{b1,t1}
\fmf{phantom}{b6,t6}
\fmf{plain,tension=2}{b2,v1}
\fmf{plain,tension=1}{v1,v2}
\fmf{plain_arrow}{v2,t2}
\fmf{plain,tension=2}{b5,v3}
\fmf{plain,tension=1}{v3,v4}
\fmf{plain_arrow}{v4,t5}
\fmffreeze
\fmf{zigzag,label=$q_1,, c_1$}{v1,v5}
\fmf{zigzag,label=$r,, i$}{v5,v3}
\fmffreeze
\fmf{plain_arrow}{t4,v6}
\fmf{plain_arrow,label=$k'_2 \qquad \qquad$,label.side=left}{v6,v5}
\fmf{plain_arrow}{v5,t3}
\fmffreeze
\fmf{zigzag,label=$q_2-r,,j$,label.side=left}{v6,v4}
\fmfdot{v1,v3,v4}
\fmfv{label=$k_1$,label.angle=90}{t3}
\fmfv{label=$-k_2$,label.angle=90}{t4}
\fmfv{label=$A'$,label.angle=90}{t2}
\fmfv{label=$B'$,label.angle=90}{t5}
\fmfv{label=$A$,label.angle=-90}{b2}
\fmfv{label=$B$,label.angle=-90}{b5}
\fmfv{decor.shape=circle,decor.filled=1,decor.size=0.05w}{v5,v6}
\fmf{phantom,label=$\textmd{\large c} $,label.side=right,label.dist=6mm}{b3,b4}
\end{fmfgraph*}
\end{fmffile}
\end{minipage}
\\[10mm]
\begin{minipage}[t]{54mm}
\setlength{\unitlength}{1.0mm}
\begin{fmffile}{fig3d}
\fontsize{8}{12pt}
\begin{fmfgraph*}(50,45)
\fmfpen{thin}
\fmfstraight
\fmfbottomn{b}{6}
\fmftopn{t}{6}
\fmfright{r}
\fmfleft{l}
\fmf{dashes}{r,l}
\fmf{phantom}{b1,t1}
\fmf{phantom}{b6,t6}
\fmf{plain}{b2,v1,v2}
\fmf{plain_arrow}{v2,t2}
\fmf{plain}{b5,v3,v4}
\fmf{plain_arrow}{v4,t5}
\fmffreeze
\fmf{zigzag,label=$r,, i$,label.dist=0.05w}{v1,v3}
\fmf{zigzag,label=$q_1-r,, j$,label.dist=0.05w,label.side=right}{v2,v5}
\fmf{zigzag,label=$\; q_2-r,, j'$,label.dist=0.05w,label.side=right}{v5,v4}
\fmffreeze
\fmf{plain_arrow}{t4,v5,t3}
\fmfv{label=$k_1$,label.angle=90}{t3}
\fmfv{label=$-k_2$,label.angle=90}{t4}
\fmfdotn{v}{4}
\fmfv{decor.shape=circle,decor.filled=1,decor.size=0.05w}{v5}
\fmfv{label=$A'$,label.angle=90}{t2}
\fmfv{label=$B'$,label.angle=90}{t5}
\fmfv{label=$A$,label.angle=-90}{b2}
\fmfv{label=$B$,label.angle=-90}{b5}
\fmf{phantom,label=$\textmd{\large d}$,label.side=right,label.dist=6mm}{b3,b4}
\end{fmfgraph*}
\end{fmffile}
\end{minipage}
\begin{minipage}[t]{54mm}
\setlength{\unitlength}{1.0mm}
\begin{fmffile}{fig3e}
\fontsize{8}{12pt}
\begin{fmfgraph*}(50,45)
\fmfpen{thin}
\fmfstraight
\fmfbottomn{b}{6}
\fmftopn{t}{6}
\fmfright{r}
\fmfleft{l}
\fmf{dashes}{r,l}
\fmf{plain}{b2,v1,v2}
\fmf{plain_arrow}{v2,t2}
\fmf{plain}{b5,v3,v4}
\fmf{plain_arrow}{v4,t5}
\fmffreeze
\fmf{zigzag,label=$\quad q_1-q_2+r,, j$}{v1,v5}
\fmf{zigzag,label=$\quad r,, j'$,label.side=left}{v5,v3}
\fmf{zigzag,label=$q_2-r,,i\qquad\qquad$,label.side=right}{v2,v4}
\fmffreeze
\fmf{plain_arrow}{t4,v6}
\fmf{plain}{v6,v5,v7}
\fmf{plain_arrow}{v7,t3}
\fmfv{label=$k_1$,label.angle=90}{t3}
\fmfv{label=$-k_2$,label.angle=90}{t4}
\fmfdotn{v}{4}
\fmfv{decor.shape=circle,decor.filled=1,decor.size=0.05w}{v5}
\fmfv{label=$A'$,label.angle=90}{t2}
\fmfv{label=$B'$,label.angle=90}{t5}
\fmfv{label=$A$,label.angle=-90}{b2}
\fmfv{label=$B$,label.angle=-90}{b5}
\fmf{phantom,label=$\textmd{\large e} $,label.side=right,label.dist=6mm}{b3,b4}
\end{fmfgraph*}
\end{fmffile}
\end{minipage}
\begin{minipage}[t]{54mm}
\setlength{\unitlength}{1.0mm}
\begin{fmffile}{fig3f}
\fontsize{8}{12pt}
\begin{fmfgraph*}(50,45)
\fmfpen{thin}
\fmfstraight
\fmfbottomn{b}{6}
\fmftopn{t}{6}
\fmfright{r}
\fmfleft{l}
\fmf{dashes}{r,l}
\fmf{plain}{b2,v1,v2}
\fmf{plain}{b5,v3,v4}
\fmf{plain_arrow}{v2,t2}
\fmf{plain_arrow}{v4,t5}
\fmffreeze
\fmf{zigzag,tension=1,label=$q_1-\tilde q_1,, i$}{v1,v6}
\fmf{zigzag,tension=2,label=$\;   r,, i'$}{v6,v3}
\fmf{zigzag,tension=1,label=$\tilde q_1 ,, j$,label.side=left}{v2,v5}
\fmf{zigzag,tension=1}{v5,v7}
\fmf{zigzag,tension=1,label=$q_2-r,,\;\;\;\;  j' \;\; \qquad \quad$,label.side=right}{v7,v4}
\fmffreeze
\fmf{curly}{v5,t3}
\fmf{curly}{v6,t4}
\fmfv{decor.shape=circle,decor.filled=1,decor.size=0.05w}{v5,v6}
\fmfdot{v1,v2,v3,v4}
\fmfv{label=$k_1$,label.angle=90}{t3}
\fmfv{label=$k_2$,label.angle=90}{t4}
\fmfv{label=$A'$,label.angle=90}{t2}
\fmfv{label=$B'$,label.angle=90}{t5}
\fmfv{label=$A$,label.angle=-90}{b2}
\fmfv{label=$B$,label.angle=-90}{b5}
\fmf{phantom,label=$\textmd{\large f}$,label.side=right,label.dist=6mm}{b3,b4}
\end{fmfgraph*}
\end{fmffile}
\end{minipage}
\\[10mm]
\begin{minipage}[c]{54mm}
\setlength{\unitlength}{1.0mm}
\begin{fmffile}{fig3g}
\fontsize{8}{12pt}
\begin{fmfgraph*}(50,45)
\fmfpen{thin}
\fmfstraight
\fmfbottomn{b}{6}
\fmftopn{t}{6}
\fmfright{r}
\fmfleft{l}
\fmf{dashes}{r,l}
\fmf{plain}{b2,v1,v2}
\fmf{plain}{b5,v3,v4}
\fmf{plain_arrow}{v2,t2}
\fmf{plain_arrow}{v4,t5}
\fmffreeze
\fmf{zigzag,tension=1,label=$q_1-\tilde q_2,, i$,label.side=left}{v1,v6}
\fmf{zigzag,tension=2,label=$q_2-r,,i'\;\;\;$,label.side=right}{v6,v3}
\fmf{zigzag,tension=1,label=$ \tilde q_2,, j$,label.side=left}{v2,v5}
\fmf{zigzag,tension=1,label=$\quad  \qquad  \qquad r,, j'$,label.side=left}{v5,v7}
\fmf{zigzag,tension=1}{v7,v4}
\fmffreeze
\fmf{curly}{v5,t3}
\fmf{curly}{v6,t4}
\fmfv{decor.shape=circle,decor.filled=1,decor.size=0.05w}{v5,v6}
\fmfdot{v1,v2,v3,v4}
\fmfv{label=$k_2$,label.angle=90}{t3}
\fmfv{label=$k_1$,label.angle=90}{t4}
\fmfv{label=$A'$,label.angle=90}{t2}
\fmfv{label=$B'$,label.angle=90}{t5}
\fmfv{label=$A$,label.angle=-90}{b2}
\fmfv{label=$B$,label.angle=-90}{b5}
\fmf{phantom,label=$\textmd{\large g} $,label.side=right,label.dist=6mm}{b3,b4}
\end{fmfgraph*}
\end{fmffile}
\end{minipage}
\end{center}
\caption{Schematic representation of the discontinuities of the
amplitude  ${\cal A}_{AB}^{A'\{P_1P_2\}B'}$: a -in the
$s_2$-channel; b -in  the $s_{21}$-channel; c -in the
$s_{22}$-channel; d -in the $s$-channel;
  e -in the $s'$-channel; f -in the $s'_1$-channel ; g -in the
  $s'_2$-channel.}
\end{figure}
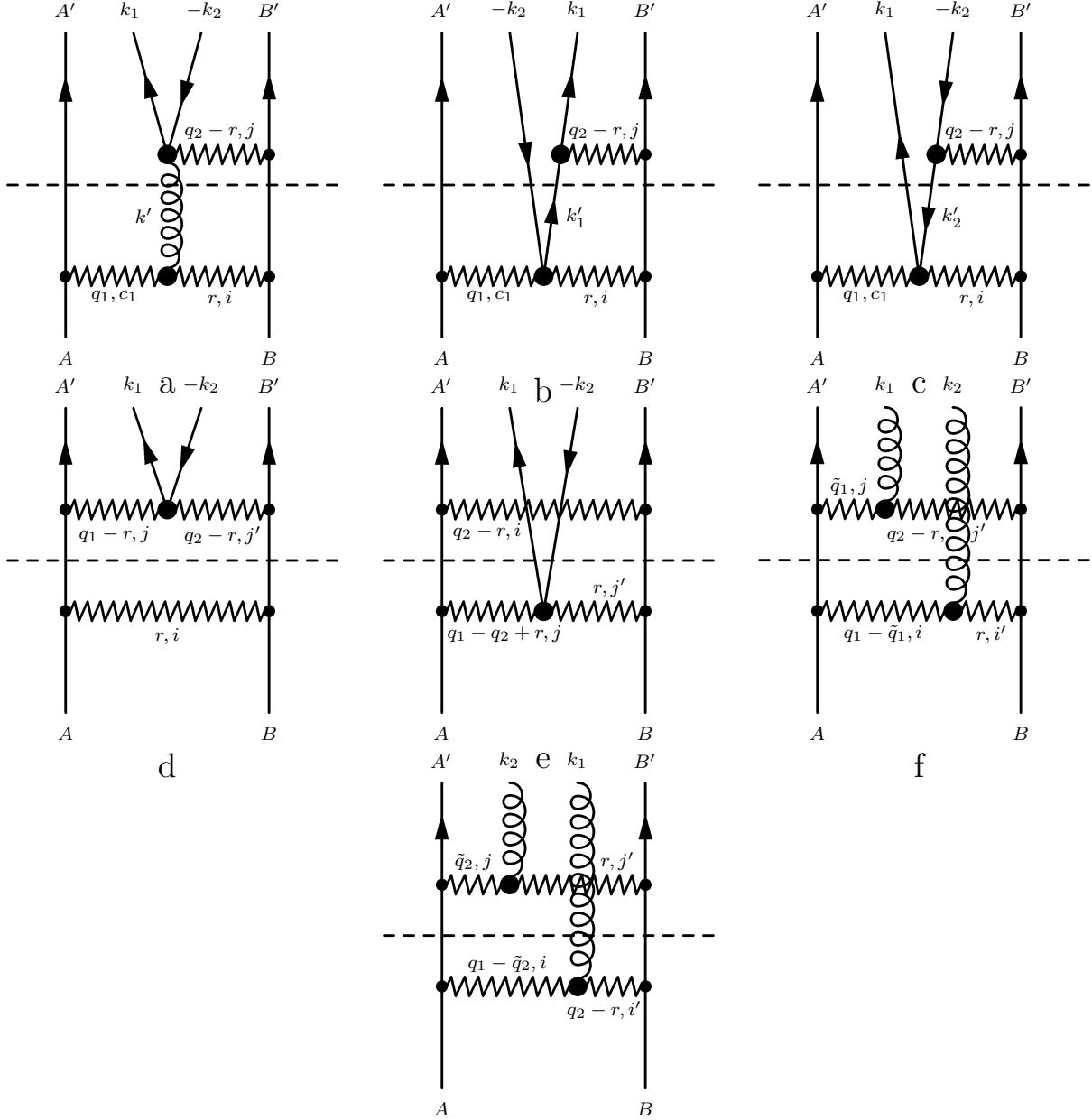
Let start with the $s_2$-channel (see Fig.~3~a):
\begin{equation}
\Im_{3a} {\cal A}_{AB}^{A'\{P_1P_2\}B'}=\frac 1{2}\sum_{\{\tilde P
\tilde B\}}\int {\cal A}^{A'\tilde P \tilde B}_{AB} {\cal
A}^{\{P_1P_2\}B'}_{\tilde P \tilde B} d\Phi _{\tilde P \tilde B}~,
\label{Im 3a}
\end{equation}
where $d\Phi _{\tilde P \tilde B}$ is given by  (\ref{rho2i}) with
the replacement $p_{P_1}\rightarrow k$.  As always, $r_\perp
=(p_{\tilde B}-p_B)_\perp$. The particle $\tilde{P}$ has to be
produced in the MRK, so that it must be a gluon. Denoting its
momentum $k'$ we have \be k'=\beta
p_1-\frac{(q_1-r)_\perp^2}{\beta s}p_2+(q_1-r)_\perp~. \ee The
possibility of the representation (\ref{Im central}) for the
imaginary part (\ref{Im 3a}) becomes evident if one  takes the
representations (\ref{z11}) and (\ref{fragment}) in the Born
approximation for the amplitudes in (\ref{Im 3a}), extracts  the
antisymmetric colour octet in the $t_2$-channel
$(t_2=(p_B-p_{B'})^2)$  by the projection operator (\ref{project})
and uses  the bootstrap property of the LO vertices (\ref{boot
lo}). For the contribution ${\cal F}_{c_1c_2}^{a}$ to ${\cal
F}^{P_1P_2}_{c_1c_2}(q_1,q_2,r_\perp)$ one obtains
\begin{equation}
{\cal F}_{c_1c_2}^{a}= if_{ijc_2}\sum_{\{G\}}\gamma^G_{c_1 i}(q_1,
q_1-k')\Gamma^{j}_{\{P_1P_2\}G}~. \label{Fa}
\end{equation}
Imaginary parts in the channels $(p_{B'}+k_1)^2$ and
$(p_{B'}+k_2)^2$ (see Fig.~3~b,c) are  found quite analogously.
For the first of them we have
\begin{equation}
\Im_{3b} {\cal A}_{AB}^{A'\{P_1P_2\}B'}=\frac 1{2}\sum_{\{\tilde P
\tilde B\}}\int {\cal A}^{A'\{\tilde P P_2\} \tilde B}_{AB} {\cal
A}^{P_1 B'}_{\tilde P \tilde B} d\Phi _{\tilde P \tilde B}~,
\label{Im 3b}
\end{equation}
where $d\Phi _{\tilde P \tilde B}$ is given now just by
(\ref{rho2i}).  Evidently, the particle $\tilde{P}$ now is of the
same kind as $P_1$. Denoting its momentum $k^{\prime}_1$ we have
\be k^{\prime}_1=\beta_1
p_1+\frac{m_1^2-(q_1-k_2-r)_\perp^2}{\beta_1
s}p_2+(q_1-k_2-r)_\perp~,  \ee where $m_1$ is its mass. The
amplitudes ${\cal A}^{A'\{\tilde P P_2\} \tilde B}_{AB}$ and
${\cal A}^{P_1 B'}_{\tilde P \tilde B}$ are given  by
(\ref{central}) and (\ref{Ael}) respectively, taken in the Born
approximation. After extraction of the antisymmetric colour octet
in the $t_2$-channel and use of (\ref{boot lo}) we come to  the
representation (\ref{Im central}) with  the contribution ${\cal
F}_{c_1c_2}^{b}$ to ${\cal F}^{P_1P_2}_{c_1c_2}(q_1,q_2,r_\perp)$
equal
\begin{equation}
{\cal F}_{c_1c_2}^{b}= if_{ijc_2}\sum_{\{\tilde P\}}\gamma^{\tilde
PP_2}_{c_1 i}(q_1, q_1-k_1^{\prime}-k_2)\Gamma^{j}_{P_1\tilde P}~.
\label{Fb}
\end{equation}
Evidently,
\begin{equation}
{\cal F}_{c_1c_2}^{c}=  {\cal F}_{c_1c_2}^{b}(P_1\leftrightarrow
P_2)~. \label{Fc}
\end{equation}
The  imaginary parts shown in Figs. 3 d-g  are calculated in a
similar way. For Fig.~3~d one has
\begin{equation}
\Im_{3d} {\cal A}_{AB}^{A'\{P_1P_2\}B'}=\frac 1{2}\sum_{\{\tilde A
\tilde B\}}\int {\cal A}^{\tilde A \tilde B}_{AB} {\cal A}^{A' \{
P_1 P_2\} B'}_{\tilde A \tilde B} d\Phi _{\tilde A \tilde B}~,
\label{Im 3d}
\end{equation}
where $d\Phi _{\tilde A \tilde B}$ is given  by (\ref{rho2}); $
r_\perp =(p_{\tilde B}-p_{B})_\perp \simeq p_{\tilde B}-p_{B}$ .
The amplitudes ${\cal A}^{\tilde A \tilde B}_{AB}$ and ${\cal
A}^{A' \{\tilde P P_2\} B'}_{\tilde P \tilde B}$ are given by the
Born terms of (\ref{central}) and (\ref{Ael}) respectively. The
difference of further  calculation from preceding ones  is that it
is necessary to apply the projection operator (\ref{project}) and
to use the bootstrap property (\ref{boot lo}) both in the $t_1$-
and $t_2$-channels. After this it becomes clear that again the
imaginary parts have the form (\ref{Im central}) with the
contribution to ${\cal F}^{P_1P_2}_{c_1c_2}(q_1,q_2,r_\perp)$
equal
\begin{equation}
{\cal F}_{c_1c_2}^{d}=\frac{g}{2}
f_{ijc_1}f_{ij'c_2}\frac{q_{1\perp}^2}{(q_1-r)_\perp^2}\gamma^{P_1P_2}_{jj'}
(q_1-r_\perp, q_2-r_\perp)~. \label{Fd}
\end{equation}
The imaginary part answering  Fig.~3~e is
\begin{equation}
\Im_{3e} {\cal A}_{AB}^{A'\{P_1P_2\}B'}=\frac 1{2}\sum_{\{\tilde A
\tilde B\}}\int {\cal A}^{\tilde A \{P_1 P_2\} \tilde B}_{AB}
{\cal A}^{A' B'}_{\tilde A \tilde B} d\Phi _{\tilde A \tilde B}~,
\label{Im 3e}
\end{equation}
where $d\Phi _{\tilde A \tilde B}$ is given now  by (\ref{rho2})
with the replacement $(p_A+p_B\rightarrow p_{A'}+p_{B'})$. It is
easy to see that the contribution of this imaginary part to ${\cal
F}^{P_1P_2}_{c_1c_2}(q_1,q_2,r_\perp)$ is obtained from  ${\cal
F}_{c_1c_2}^{d}$ by the substitution $r\leftrightarrow q_2-r$.
Since the integration measure in (\ref{Im central}) is invariant
under this substitution, we can put
\begin{equation}
{\cal F}_{c_1c_2}^{e}={\cal F}_{c_1c_2}^{d}~. \label{Fe}
\end{equation}
At last, Figs. 3 f,g appear only in the case when the particles
$P_1$ and $P_2$ are gluons. The imaginary part answering Fig.~3~f
is
\begin{equation}
\Im_{3f} {\cal A}_{AB}^{A'\{P_1P_2\}B'}=\frac 1{2}\sum_{\{\tilde A
\tilde B\}}\int {\cal A}^{\tilde A P_2 \tilde B}_{AB} {\cal A}^{A'
 P_1 B'}_{\tilde A  \tilde B} d\Phi _{\tilde A \tilde B}~.
\label{Im 3f}
\end{equation}
The amplitudes entering in (\ref{Im 3f}) are given by (\ref{z11})
with $n=1$ in the Born approximation. Again applying  the
projection operator (\ref{project}) and using the bootstrap
property (\ref{boot lo}) in the $t_1$- and $t_2$-channels we
obtain 
\begin{equation}
{\cal F}_{c_1c_2}^{f}= \frac{g}{2}
f_{i'j'c_2}f_{ijc_1}
\frac{q_{1\perp}^2}{\tilde q^2_{1\perp}
(q_1-\tilde q_1)_\perp^2}\gamma^{P_2}_{ii'}(q_1-\tilde q_1,
q_1-\tilde{q_1}-k_2) \gamma^{P_1}_{jj'}(\tilde{q_1},\tilde{q_1}-k_1), 
\label{Ff}
\end{equation}
where $\tilde q_1=\beta_1 p_1 +(k_1+q_2-r)_\perp$.
Evidently,
\begin{equation}
{\cal F}_{c_1c_2}^{g}=  {\cal F}_{c_1c_2}^{f}(P_1\leftrightarrow
P_2)~. \label{Fg}
\end{equation}
Note that ${\cal F}_{c_1c_2}^{f}$ is invariant under simultaneous
substitution $P_1\leftrightarrow P_2$  (that means, in particular,
$k_1\leftrightarrow k_2$) and $r_\perp\leftrightarrow
(q_2-r)_\perp $. The last substitution can be considered as  the
redefinition of $r_\perp $. Since the integration measure in
(\ref{Im central}) is invariant under this redefinition, we can
make
\begin{equation}
{\cal F}_{c_1c_2}^{g}=  {\cal F}_{c_1c_2}^{f}~. \label{Fg1}
\end{equation}
Therefore, we have
\begin{equation}
{\cal F}^{P_1P_2}_{c_1c_2}(q_1,q_2,r)={\cal F}_{c_1c_2}^{a}+{\cal
F}_{c_1c_2}^{b}+{\cal F}_{c_1c_2}^{c}+2{\cal
F}_{c_1c_2}^{d}+2{\cal F}_{c_1c_2}^{f}~, \label{F}
\end{equation}
where the terms in the R.H.S. are given respectively by Eqs.
(\ref{Fa}), (\ref{Fb}), (\ref{Fc}), (\ref{Fd}) and (\ref{Ff}).

As it was discussed earlier, the terms with $\ln s_2$ in the real
part of the amplitude ${\cal A}_{AB}^{A'\{P_1P_2\}B'}$ are
obtained from (\ref{Im central}) by the substitution (\ref{sub s})
with $s_2$ instead of $s$. Comparing the obtained result with
(\ref{central}) with account of (\ref{omega_1}), we see that the
one-loop correction calculated above is compatible with the
Reggeized form (\ref{central}) only if
\begin{equation}
\int \frac{d^{D-2}r_{\perp}}{r_{\perp} ^2(q_2-r)^{2}_\perp} {\cal F}^{P_1P_2}_{c_1c_2}(q_1,q_2,r_\perp)
=\frac{gN_c}{2}\gamma_{c_1c_2}^{\{P_1P_2\}}(q_1,q_2) \int
\frac{d^{D-2}r_{\perp}}{r_{\perp} ^2(q_2-r)^{2}_\perp}~.
\label{boot central}
\end{equation}
Eq.(\ref{boot central}) gives  the bootstrap conditions for the
vertices of  pair production in  Reggeon-Reggeon collisions. They
are verified in the next subsections.

\subsection{Quark-antiquark production}

For simplicity, we discuss below the case of the massless quarks,
although the massive case can be considered quite analogously.

\subsubsection*{Denotations}

Remind that $k_1$ and $k_2$ are the quark and antiquark momenta
respectively;
$$ k_i=\beta_i p_1 +\alpha_i p_2 +k_{i\perp}~,
\;\;i=1,2~,\;\;s\alpha_i\beta_i =-k_{i\perp}^2=\vec k_i^{~2}~; $$
\be \beta_i=x_i \beta~, \;\; \beta=\beta_1+\beta_2~;
\;\;k=k_1+k_2=q_1-q_2~, \label{kin1} \ee and we can put \be
q_1=\beta p_1 +q_{1\perp}~, \;\; q_2=-\alpha p_2+q_{2\perp}~, \;\;
\beta =\beta_1+\beta_2~, \alpha =\alpha_1+\alpha_2~. \ee We use
also $$ k'=\beta p_1-\frac{(q_1-r)_\perp^2}{\beta
s}p_2+(q_1-r)_\perp~, \;\; k^{\prime}_1=\beta_1
p_1-\frac{(q_1-k_2-r)_\perp^2}{\beta_1 s}p_2+(q_1-k_2-r)_\perp~,
$$ \be k^{\prime}_2=\beta_2
p_1-\frac{(q_1-k_1-r)_\perp^2}{\beta_2
s}p_2+(q_1-k_1-r)_\perp~.\label{kin2} \ee The function ${\cal
F}^{P_1P_2}_{c_1c_2}(q_1,q_2,r_\perp)$ in (\ref{boot central}) is
expressed in terms of the Reggeon vertices defined in
(\ref{Gamma_Q}), (\ref{Gamma_bar Q}), (\ref{gammaRRG}),
(\ref{Gamma gqq}) and  the effective vertex of  quark-antiquark
production in  Reggeon-Reggeon collisions. The last vertex was
found in ~\cite{FL96} and has the form
\begin{equation}
\gamma _{c_1c_2}^{Q{\bar Q}}(q_1,q_2)= \frac 12g^2\bar u(k_1)
\left[ t^{c_1}t^{c_2}a(q_1;k_1,k_2)\,-t^{c_2}\,t^{c_1}\overline{
a(q_1;k_2,k_1)}\right] v(k_2)~,  \label{gammaqbarq}
\end{equation}
where $a(q_1;k_1,k_2)$ and $\overline{a(q_1;k_2,k_1)}$ can be
written   \cite{FFFK} in the following way:
\begin{equation}
a(q_1;k_1,k_2)=\frac{4\not{p}_1\not{Q}_1\not{p}_2}{s\tilde{t}_1}-\frac{1}{k^2}\not{\Gamma}\,,
\;\overline{a(q_1;k_2,k_1)}=\frac{4\not{p}_2\not{Q}_2\not{p}_1}
{s\tilde{t}_2}-\frac{1}{k^2}\not{\Gamma}~,
\end{equation}
with
\[
~\tilde t_1=(q_1-k_1)^2~,~~~\tilde t_2=(q_1-k_2)^2~,~~~
Q_1=q_{1\perp }-k_{1\perp }~,~~~Q_2=q_{1\perp }-k_{2\perp }~,
\]
\begin{equation}
\Gamma =2\left[ (q_1+q_2)_{\perp }-\beta p_A\left( 1-2\frac{\vec
q_1^{\:2}}{ s\alpha \beta }\right) +\alpha p_B\left( 1-2\frac{\vec
q_2^{\:2}}{s\alpha \beta }\right) \right] ~.
\end{equation}
Further for denominators in the Reggeon vertices we use
denotations $D(p,q)$ and $d(p,q)$: \be
D(p,q)=x_1p_{\bot}^2+x_2q_{\bot}^2~,\;\;
d(p,q)=(x_1p_{\bot}-x_2q_{\bot})^2;\;\;
D(p,q)=d(p,q)+x_1x_2(p_{\bot}+q_{\bot})^2. \label{Dandd} \ee
Seeing that for arbitrary $p_\perp$
\begin{equation}
\bar u(k_1)p\!\!\!/_{\perp} v(k_2)=\bar
u(k_1)\frac{p\!\!\!/_{2}}{s\beta}
\left(\frac{k\!\!\!/_{1\perp}p\!\!\!/_{\perp}}{x_1}+\frac{p\!\!\!/_{\perp}k\!\!\!/_{2\perp}}{x_2}\right)
v(k_2)~,
\end{equation}
we can present  $a(q_1; k_1,k_2)$ and $\overline{a(q_1; k_2,k_1)}$
 as \be a(q_1;
k_1,k_2)=\frac{4}{s\beta}p\!\!\!/_{2} b(q_1; k_1,k_2)~, \;\;
\overline{a(q_1; k_2,k_1)}= \frac{4}{s\beta}p\!\!\!/_{2}
\overline{b(q_1; k_2,k_1)}~,\label{atob} \ee where
\[
b(q_1; k_1,k_2)=\frac{k\!\!\!/_{1\perp}
(k\!\!\!/_{1\perp}-q\!\!\!/_{1\perp})}{D(k_1-q_1,k_1)}
-\frac{x_1x_2}{d(k_2,k_1)}\left(\frac{q_{1\perp}^2k\!\!\!/_{1\perp}
k\!\!\!/_{2\perp}}{D(k_2,k_1)}\right.
\]
\[
\left.
-\frac{k\!\!\!/_{1\perp}q\!\!\!/_{1\perp}}{x_1}-\frac{q\!\!\!/_{1\perp}k\!\!\!/_{2\perp}}{x_2}
-q_{1\perp}^2+2(q_{1\perp}(k_1+k_2)_{\perp})\right)-1~,
\]
\[
\overline{b(q_1; k_2,k_1)}=\frac{(k\!\!\!/_{2\perp}
-q\!\!\!/_{1\perp})k\!\!\!/_{2\perp} } {D(k_2,k_2-q_1)}
-\frac{x_1x_2}{d(k_2,k_1)}\left(\frac{q_{1\perp}^2k\!\!\!/_{1\perp}
k\!\!\!/_{2\perp}}{D(k_2,k_1)}\right.
\]
\be
 \left.
-\frac{k\!\!\!/_{1\perp}q\!\!\!/_{1\perp}}{x_1}-\frac{q\!\!\!/_{1\perp}
k\!\!\!/_{2\perp}}{x_2}
-q_{1\perp}^2+2(q_{1\perp}(k_1+k_2)_{\perp})\right)-1~.\label{bbarb}
\ee This form of $a(q_1; k_1,k_2)$ and $\overline{a(q_1;
k_2,k_1)}$ permits to perform quite readily  the summation over
spin projections $\lambda$ of intermediate quarks and antiquarks
in the contributions ${\cal F}_{c_1c_2}^{b}$ and ${\cal
F}_{c_1c_2}^{c}$ to  ${\cal
F}^{P_1P_2}_{c_1c_2}(q_1,q_2,r_\perp)$; for example: \be
\sum_{\lambda}\bar u(k_1)\frac{p\!\!\!/_2}{\beta_1 s} u^\lambda
(k'_1) \bar u^\lambda (k'_1) a(q_1;k'_1,k_2) v(k_2)=\bar u(k_1)
a(q_1,k'_1,k_2) v(k_2)~. \label{spinsum} \ee

\subsubsection*{Independent colour structures}

It is easy to calculate the number of independent colour
structures for  production of  a $q\bar q$ pair by two Reggeized
gluons. Indeed, the pair can be either in a colour singlet, or in a
colour octet state. Due to the colour symmetry each of these state
can be produced only by the same state of two Reggeized gluons,
which are colour octets. Since there is one singlet and two octets
(symmetric and antisymmetric)  in  decomposition of product of two
octets into  irreducible representations, the number of
independent colour structures is three. Their choice is not
unique. We accept the following one:
\begin{equation}
{\cal R}_1^{c_1c_2}=\frac{1}{N_c}f^{c_1\,i\,a}f^{c_2\,i\,b}(t^a\,t^b+t^b\,t^a)~,\;\;
{\cal R}_2^{c_1c_2}=i\,f^{c_1\,c_2\,i}\,t^i~, \;\;
{\cal R}_3^{c_1c_2}=t^{c_1}\,t^{c_2}+t^{c_2}\,t^{c_1}~.\label{eq:colorstr}
\end{equation}
From the equality
\be
t^a\,t^b=\frac{1}{2N_c}\delta^{a\,b}+\frac{1}{2}d^{a\,b\,c}\:t^c%
+\frac{1}{2}i\,f^{a\,b\,c}\:t^c \ee it is seen that the first and the third structures contain a singlet and a symmetric octet,
whereas the second structure contains only an antisymmetric octet.

\subsubsection*{Representation of ${\cal F}^{P_1P_2}_{c_1c_2}(q_1,q_2,r)$}

Using these colour structure we can represent each of the
contribution ${\cal F}_{c_1c_2}^{i}$ entering in  ${\cal
F}^{P_1P_2}_{c_1c_2}(q_1,q_2,r)$ (\ref{F}) in the form \be {\cal
F}_{c_1c_2}^{i}=\frac{g^3N_c}{s \beta}\bar u (k_1){p\!\!\!/_{2}}
\sum_{n=1}^{n=3}{\cal R}_n^{c_1c_2}{\cal L}^i_n v(k_2)~.
\label{Fi} \ee It is not difficult to find all ${\cal L}^i_n$ from
the equations presented above.

From (\ref{Fa}), using the Reggeon-Reggeon-gluon (RRG) vertex
(\ref{gammaRRG}) in the gauge (\ref{vectorC1}) and the vertex for
$q\bar q$ production in the fragmentation region (\ref{Gamma gqq})
we obtain:
\[
{\cal L}^a_1=
\frac{x_1\not{q}_{1\bot}(\not{k}_{1\bot}-x_1\not{k}'_{\bot})-
x_2(\not{k}_{1\bot}-x_1\not{k}'_{\bot})\not{q}_{1\bot}}
{d(k'_2,k_1)}+\frac {
x_2(x_2\not{k}'_{\bot}-\not{k}_{2\bot})\not{q}_{1\bot}-
x_1\not{q}_{1\bot}(x_2\not{k}'_{\bot}-\not{k}_{2\bot})}
{d(k_2,k'_1)}
\]
\begin{equation}
 +\frac{q_{1\bot}^2\bigl(x_2(\not{k}_{1\bot}-x_1\not{k}'_{\bot})\not{k}'_{\bot}
-x_1\not{k}'_{\bot}(\not{k}_{1\bot}-x_1\not{k}'_{\bot})\bigr)}
{k_{\bot}^{'2}d(k'_2,k_1)} +\frac{q_{1\bot}^2
\bigl(x_1\not{k}'_{\bot}(x_2\not{k}'_{\bot}-\not{k}_{2\bot})
-x_2(x_2\not{k}'_{\bot}-\not{k}_{2\bot})\not{k}'_{\bot}\bigr)}{k_{\bot}^{'2}d(k_2,k'_1)}~;
\label{La1}
\end{equation}
\[
{\cal L}^a_2= \frac {
    x_2(x_2\not{k}'_{\bot}-\not{k}_{2\bot})\not{q}_{1\bot}-
    x_1\not{q}_{1\bot}(x_2\not{k}'_{\bot}-\not{k}_{2\bot})}
{d(k_2,k'_1)}- \frac {
    x_1\not{q}_{1\bot}(\not{k}_{1\bot}-x_1\not{k}'_{\bot})-
    x_2(\not{k}_{1\bot}-x_1\not{k}'_{\bot})\not{q}_{1\bot}}
{d(k'_2,k_1)}
\]
\[
-\frac{q_{1\bot}^2\bigl(x_2(\not{k}_{1\bot}-x_1\not{k}'_{\bot})\not{k}'_{\bot}
-x_1\not{k}'_{\bot}(\not{k}_{1\bot}-x_1\not{k}'_{\bot})\bigr)}
{k_{\bot}^{'2}d(k'_2,k_1)}
+\frac{q_{1\bot}^2\bigl(x_1\not{k}'_{\bot}(x_2\not{k}'_{\bot}-\not{k}_{2\bot})
-x_2(x_2\not{k}'_{\bot}-\not{k}_{2\bot})\not{k}'_{\bot}\bigr)}
{k_{\bot}^{'2}d(k_2,k'_1)}
\]
\[
+2\frac
{x_1\not{q}_{1\bot}(x_2\not{k}_{1\bot}-x_1\not{k}_{2\bot})-
x_2(x_2\not{k}_{1\bot}-x_1\not{k}_{2\bot})\not{q}_{1\bot}}
{d(k_2,k_1)}
\]
\begin{equation}
+2\frac{q_{1\bot}^2}{k_{\bot}^{'2}d(k_2,k_1)}\bigl(
x_2(x_2\not{k}_{1\bot}-x_1\not{k}_{2\bot})\not{k}'_{\bot}-
x_1\not{k}'_{\bot}(x_2\not{k}_{1\bot}-x_1\not{k}_{2\bot}) \bigr)~;
\label{La2}
\end{equation}
\begin{equation}
{\cal L}^a_3=0~. \label{La3}
\end{equation}
In the case of $q\bar q$ production  the particle $\tilde P$ in
the sum (\ref{Fb}) must be a quark with momentum $k^{\prime}_1$.
Taking the representation  (\ref{gammaqbarq}), (\ref{atob}) for
the vertex of quark-antiquark production in  Reggeon-Reggeon
collisions, (\ref{Gamma_Q}) for the Quark-Quark-Reggeon vertex and
summing over  spin projections according to (\ref{spinsum}), we
have:
\begin{equation}
{\cal L}^b_1=- b(q_1;k'_1,k_2)~,\;\; {\cal L}^b_2=-\frac{1}{2}
\overline{b(q_1;k_2,k'_1)}~, \;\; {\cal L}^b_3=
\frac{1}{2}(b(q_1;k'_1,k_2)-\overline{b(q_1;k_2,k'_1)})~.
\label{Lb123}
\end{equation}
Quite analogously we obtain
\begin{equation}
{\cal L}^c_1=-\overline{b(q_1;k'_2,k_1)}~, \;\; {\cal
L}^c_2=\frac{1}{2} b(q_1;k_1,k'_2)~,\;\; {\cal L}^c_3=-
\frac{1}{2}(b(q_1;k_1,k'_2)-\overline{b(q_1;k'_2,k_1)})~.
\label{Lc123}
\end{equation}
The functions $b(q_1;k_1,k_2)$ and $\overline{ b(q_1;k_2,k_1)}$
are defined in  (\ref{bbarb}).

The quantities ${\cal L}^c_n$ are easily obtained from (\ref{Fd})
with account of the representation (\ref{gammaqbarq}),
(\ref{atob}) and are equal
\begin{equation}
{\cal L}^d_1=\frac{q_{1\bot}^2}{k_{\bot}^2}\bigl( b(q_1-r;
k_1,k_2)-\overline{b(q_1-r;k_2,k_1)}\bigr)~,\label{Ld1}
\end{equation}
\begin{equation}
{\cal L}^d_2=-\frac{q_{1\bot}^2}{k_{\bot}^2}\bigl(
b(q_1-r;k_1,k_2)+\overline{b(q_1-r;k_2,k_1)}\bigr)~, \label{Ld2}
\end{equation}
\begin{equation}
{\cal L}^d_3=0~. \label{Ld3}
\end{equation}
Since in the case of  $q\bar q$ production the diargams
Fig.~3~f,g can not contribute,  Eqs. (\ref{Fi})-(\ref{Ld3})
together with (\ref{F}) determine the L.H.S. of the bootstrap
equation (\ref{boot central}). Using (\ref{gammaqbarq}),
(\ref{atob}) and (\ref{bbarb}) we can present  the R.H.S. in the
form \be g\frac{N_c}{2}\gamma^{Q\bar
Q}_{c_1c_2}(q_1,q_2)=\frac{g^3N_c}{s \beta}\bar u
(k_1){p\!\!\!/_{2}} \sum_{n=1}^{n=3}{\cal R}_n^{c_1c_2}{\cal L}_n
v(k_2)~,  \label{R} \ee where
\begin{equation}
{\cal L}_1=0~, \label{L1}
\end{equation}
\begin{equation}
{\cal L}_2=-\frac{1}{2}\bigl( b(q_1;
k_1,k_2)+\overline{b(q_1;k_2,k_1)}\bigr)~,\label{L2}
\end{equation}
\begin{equation}
{\cal L}_3=\frac{1}{2}\bigl(
b(q_1-r;k_1,k_2)-\overline{b(q_1-r;k_2,k_1)}\bigr)~.  \label{L3}
\end{equation}

\subsubsection*{Verification of the bootstrap equation}

We have to compare the coefficients in the decomposition into the
colour structures ${\cal R}_n^{c_1c_2}$  in the left and right
parts of the bootstrap equation (\ref{boot central}). Let start
with ${\cal R}_1^{c_1c_2}$. Consider  sum of ${\cal L}^i_1$. Note
that due to the symmetry of the integration measure in (\ref{boot
central}) under the substitution
$r_{\bot}\,\rightarrow\,(q_{2\bot}-r_{\bot})$ we can make this
substitution in separate terms in ${\cal L}^i_1$. Doing it in the
terms with  the denominator  $D(k'_2,k'_2-q_1)$ permits to
convert them in terms with the denominator $D(k'_1-q_1,k'_1)$.
After that, using the decompositions
\begin{equation}\label{eq:denom1}
\frac{x_1x_2}{d(k_2,k'_1)D(k_2,k'_1)}=\frac{1}{k_{\bot}^{'2}}\biggl(\frac{1}{d(k_2,k'_1)}
-\frac{1}{D(k_2,k'_1)}\biggr)~,
\end{equation}
\begin{equation}\label{eq:denom2}
\frac{x_1x_2}{d(k'_2,k_1)D(k'_2,k_1)}=\frac{1}{k_{\bot}^{'2}}\biggl(\frac{1}{d(k'_2,k_1)}
-\frac{1}{D(k'_2,k_1)}\biggr)~,
\end{equation}
it is easy to see, that the terms with the denominators
\be
D(k'_2,k_1)~,\;\;D(k_2,k'_1)~,\;\;D(k'_1,k'_1-q_1)
\ee are
cancelled and we obtain for the sum of ${\cal L}^i_1$:
\begin{equation}\label{eq:vcol1}
\frac{x_1x_2q_{1\bot}^2}{d(k'_2,k_1)}-\frac{x_1x_2q_{1\bot}^2}{d(k_2,k'_1)}+
\frac{q_{1\bot}^2\bigl(d(k'_2,k_1)-x_1x_2k_{\bot}^{'2}\bigr)}{k_{\bot}^{'2}d(k'_2,k_1)}
+\frac{q_{1\bot}^2\bigl(x_1x_2k_{\bot}^{'2}-d(k_2,k'_1)\bigr)}{k_{\bot}^{'2}d(k_2,k'_1)}=0~,
\end{equation}
as it must be, since the structure  ${\cal R}_1^{c_1c_2}$ is
absent in the R.H.S. of the bootstrap equation.

Turn to the colour structure  ${\cal R}_2^{c_1c_2}$. Using
(\ref{eq:denom1}),(\ref{eq:denom2}) we obtain from the sum of
${\cal L}^i_2$:
\[
-\frac{x_1x_2q_{1\bot}^2}{d(k'_2,k_1)}-\frac{x_1x_2q_{1\bot}^2}{d(k_2,k'_1)}+
\frac{2}{d(k_2,k_1)}\bigl[
2x_1x_2\bigl(q_{1\bot}(k_{1\bot}+k_{2\bot})\bigr)-x_1\not{q}_{1\bot}\not{k}_{2\bot}
-x_2\not{k}_{1\bot}\not{q}_{1\bot}
\bigr]
\]
\[
+\frac{q_{1\bot}^2}{k_{\bot}^{'2}d(k'_2,k_1)}\bigl(x_1x_2k_{\bot}^{'2}-
d(k'_2,k_1)\bigr)
+\frac{q_{1\bot}^2}{k_{\bot}^{'2}d(k_2,k'_1)}\bigl(x_1x_2k_{\bot}^{'2}
-d(k_2,k'_1)\bigr)
\]
\[
-\frac{q_{1\bot}^2}{k_{\bot}^{'2}d(k_2,k_1)}(2x_1x_2k_{\bot}^{'2})
+2\frac{q_{1\bot}^2}{k_{\bot}^{'2}}
+\frac{2x_1x_2q_{1\bot}^2\not{k}_{1\bot}\not{k}_{2\bot}}{d(k_2,k_1)D(k_2,k_1)}
\]
\begin{equation}
-\frac{(\not{k}_{2\bot}-\not{q}_{1\bot})\not{k}_{2\bot}}{D(k_2,k_2-q_1)}-
\frac{\not{k}_{1\bot}(\not{k}_{1\bot}-\not{q}_{1\bot})}{D(k_1-q_1,k_1)}+2~.
\end{equation}
One can readily see that the terms depending on $r_\perp$ cancel
each other with the result:
\[
-\frac{1}{D(k_1-q_1,k_1)}\bigl(\not{k}_{1\bot}(\not{k}_{1\bot}
-\not{q}_{1\bot})\bigr)
-\frac{1}{D(k_2,k_2-q_1)}\bigl((\not{k}_{2\bot}-\not{q}_{1\bot})
\not{k}_{2\bot}\bigr)
\]
\[
-2\frac{1}{d(k_2,k_1)}\bigl[x_2\not{k}_{1\bot}\not{q}_{1\bot}
+x_1\not{q}_{1\bot}\not{k}_{2\bot}+x_1x_2q_{1\bot}^2-
2x_1x_2\bigl(q_{1\bot}(k_{1\bot}+k_{2\bot})\bigr)\bigr]
\]
\begin{equation}
+\frac{x_1x_2}{d(k_2,k_1)D(k_2,k_1)}(2\not{k}_{1\bot}\not{k}_{2\bot}q_{1\bot}^2)+2~.
\end{equation}
It is just ${\cal L}_2$, so that for the colour structure ${\cal
R}_3^{c_1c_2}$ the bootstrap equation is satisfied.

Al last, consider the colour structure ${\cal R}_3^{c_1c_2}$. For
the sum of ${\cal L}^i_3$ we have
\[
\frac{\not{k}'_{1\bot}(\not{k}'_{1\bot}-\not{q}_{1\bot})}{D(k'_1-q_1,k'_1)}-
\frac{(\not{k}_{2\bot}-\not{q}_{1\bot})\not{k}_{2\bot}}{D(k_2,k_2-q_1)}
+\frac{\not{k}_{1\bot}(\not{k}_{1\bot}-\not{q}_{1\bot})}{D(k_1-q_1,k_1)}-
\frac{(\not{k}'_{2\bot}-\not{q}_{1\bot})\not{k}'_{2\bot}}{D(k'_2,k'_2-q_1)}
\]
\begin{equation}
=\frac{\not{k}_{1\bot}(\not{k}_{1\bot}-\not{q}_{1\bot})}{D(k_1-q_1,k_1)}-
\frac{(\not{k}_{2\bot}-\not{q}_{1\bot})\not{k}_{2\bot}}{D(k_2,k_2-q_1)}~,
\end{equation}
that is exactly  ${\cal L}_3$.

So, the bootstrap equation for  $q\bar q$ production is satisfied.
\subsection{Two-gluon production}
\subsubsection*{Denotations}
In the case of two-gluon production  Eqs.
(\ref{kin1})-(\ref{kin2}) are applied as before; but now $k_1$ and
$k_2$ are the gluon momenta.   The effective vertex of two-gluon
production in Reggeon-Reggeon collisions in a gauge invariant form
was obtained in \cite{FaLi89}. In the lightcone gauge
(\ref{axial2}) for both gluons the vertex takes the form:
\[\gamma_{i j}^{G_1G_2}(q_1,q_2)=4g^2(e_{1\perp}^*)_{\alpha}
(e_{2\perp}^*)_{\beta}
\]
\begin{equation}
\times\left[\left(T^{i_1} T^{i_2}\right)_{i j} b^{\alpha\beta}
(q_1;k_1,k_2) +\left(T^{i_2} T^{i_1}\right)_{i j} b^{\beta\alpha}
(q_1;k_2,k_1)\right]~, \label{gamma_GG}
\end{equation}
where $e_{1,2}$ are the polarization vectors of the produced
gluons, $i_{1,2}$ are their colour indices, $i,j$ are the colour
indices of the Reggeons with momenta $q_1$ and $q_2$
correspondingly, and
\[
b^{\alpha\beta}(q_1;k_1,k_2)
=\frac{1}{2}g^{\alpha\beta}_{\perp}\left[
\frac{x_1x_2}{d(k_2,k_1)}(2q_{1\perp}(x_1
k_2-x_2k_1)_{\perp}+q^2_{1\perp}(x_2-\frac{x_1k^2_{2\perp}}{D(k_2,k_1)}))
\right.
\]
\[
\left.-x_2(1-\frac{k^2_{1\perp}}{D(q_1-k_1,k_1)})\right]-\frac{x_2k^{\alpha}_{1\perp}q^{\beta}_{1\perp}-x_1q^{\alpha}_{1\perp}(q_1-k_1)^{\beta}_{\perp}}{D(q_1-k_1,k_1)}-\frac{
x_1q^2_{1\perp}k^{\alpha}_{1\perp}(q_1-k_1)^{\beta} }{k^2_{1\perp}
D(q_1-k_1,k_1)}
\]
\[
-\frac{x_1q^{\alpha}_{1\perp}(x_1k_2-x_2k_1)^{\beta}_{\perp}
+x_2q^{\beta}_{1\perp}(x_1k_2-x_2k_1)^{\alpha}_{\perp}}{d(k_2,k_1)}
+\frac{x_1q^2_{1\perp}k^{\alpha}_{1\perp}k^{\beta}_{2\perp}}{k^2_{1\perp}D(k_2,k_1)}
\]
\begin{equation}
+\frac{x_1x_2q^{2}_{1\perp}}{d(k_2,k_1)D(k_2,k_1)}\Big[(x_1k_2-x_2k_1)^{\alpha}_{\perp}
k^{\beta}_{2\perp}+k^{\alpha}_{1\perp}(x_1k_2-x_2k_1)^{\beta}_{\perp}\Big]~.
\label{b}
\end{equation}
Here we use the denotations (\ref{Dandd}). Note that one can come
to (\ref{b}) starting from the vertex  in the gauge
$e(k_1)p_1=0~,\;e(k_2)p_2=0 $ \cite{FKL98}. Our
$b^{\alpha\beta}(q_1;k_1,k_2)$ can be obtained from
$c^{\alpha\beta}(k_1,k_2)$ defined in~\cite{FKL98} be the gauge
transformation
\begin{equation}
b^{\alpha\beta}(q_1;k_1,k_2) = \left(g_{\perp}^{\alpha\gamma} -
2\frac{k_{1\perp}^{\alpha}k_{1\perp}^{\gamma}}{k_{1\perp}^2}\right)
c_{\gamma}^{\; \beta}(k_1,k_2). \label{z27}
\end{equation}

\subsubsection*{Independent colour structures}
Contrary to the case of $q\bar q$ production where ${\cal
F}^{P_1P_2}_{ij}(q_1,q_2,r)$ (\ref{F}) has the most general form
in colour space, here not all admitted colour structures are
present. The number of all independent structures is readily
calculated. Indeed, decomposition of product of two octets
($8\otimes8=1\oplus8_{s}\oplus8_{a}\oplus10\oplus10^{*}\oplus27$)
contains  5  different irreducible representations, one of which
enters two times. Such decomposition is valid for two Reggeons as
well as for two gluons. Therefore, total number of admitted
independent colour structures is 8. It occurs that only three of
them  enter in ${\cal F}^{G_1G_2}_{ij}$. Actually it is
predictable and is related to specific colour structures   of the
effective vertices for one-gluon (\ref{gammaRRG}) and two-gluon
production (\ref{Gamma ggg }),(\ref{gamma_GG}). These vertices are
expressed in terms of  the colour group generators in the adjoint
representation. From properties of these generators it follows
that only three independent tensors with four indices can be built
from them. Of course, their choice is not unique. We accept the
following:
\begin{equation}
{\cal R}^{i_1 i_2}_{(1)i
j}=\frac{2}{N_c}Tr(T^{i}T^{j}T^{i_2}T^{i_1}), \;\; {\cal R}^{i_1
i_2}_{(2)i  j}=T^{i_1}_{i l}T^{i_2}_{l j}, \;\; {\cal R}^{i_1
i_2}_{(3)i  j}=T^{i_2}_{i l}T^{i_1}_{l j}.
\end{equation}
It seems that our choice is the most appropriate, i.e. the
coefficients with which these tensors enter in  ${\cal
F}^{G_1G_2}_{ij}$ are the least cumbersome.

Let us present each of the contributions  ${\cal F}_{ij}^{m}$
entering in ${\cal F}^{G_1G_2}_{ij}(q_1,q_2,r)$ (\ref{F})  in the
form \be {\cal F}_{ij}^{m}=2g^3N_c\sum_{n=1}^{n=3}{\cal R}_{(n) i
j }^{i_1i_2}(e_{1\perp}^*)_{\alpha} (e_{2\perp}^*)_{\beta}{\cal
L}^{\alpha\beta}_{mn}~. \label{Fgi} \ee Writing in the same form
the right part of (\ref{boot central}) \be
g\frac{N_c}{2}\gamma^{G_{1} G_{2}}_{ij}(q_1,q_2)=2g^3N_c
(e_{1\perp}^*)_{\alpha}
(e_{2\perp}^*)_{\beta}\sum_{n=1}^{n=3}{\cal R}_{(n) i
j}^{i_1i_2}{\cal L}^{\alpha\beta}_{n} ~, \label{Rg} \ee we have
from (\ref{gamma_GG})
\begin{equation}
{\cal L}^{\alpha\beta}_{1}=0~,\;\; {\cal
L}^{\alpha\beta}_{2}=b^{\alpha\beta}(q_1;k_1,k_2)~,\;\; {\cal
L}^{\alpha\beta}_{3}=b^{\beta\alpha}(q_1;k_2,k_1)~.\label{l123}
\end{equation}
The coefficients ${\cal L}^{\alpha\beta}_{mn}$ in (\ref{Fgi}) are
found by straightforward calculation using the vertices
(\ref{Gamma_G}), (\ref{gammaRRG}), (\ref{Gamma ggg }) and
(\ref{gamma_GG}). With account of (\ref{F}) the bootstrap
condition (\ref{boot central}) requires
\begin{equation}
\int \frac{d^{D-2}r_{\perp}}{r_{\perp}^2(q_2-r)_\perp^{2}}
\left({\cal L}^{\alpha\beta}_{an}+{\cal
L}^{\alpha\beta}_{bn}+{\cal L}^{\alpha\beta}_{cn}+2{\cal
L}^{\alpha\beta}_{dn}+2{\cal L}^{\alpha\beta}_{fn}\right)={\cal
L}^{\alpha\beta}_{n}\int \frac{d^{D-2}r_{\perp}}{r_{\perp}
^2(q_2-r)^{2}_\perp}~, \label{boot L}
\end{equation}
for each $n$.

\subsubsection*{Verification of the bootstrap equation}

For $n=1$ we obtain:
\[
 {\cal L}^{\alpha\beta}_{a1}=g^{\alpha \beta}_{\perp}x_1
x_2\Biggr[ \frac{(k_1-x_1k')_{\perp}Q_{\perp}}{d(k'_2,k_1)}+
\frac{(k_2-x_2k')_{\perp}Q_{\perp}}{d(k_2,k'_1)} \Biggr]
\]
\be
 -\frac{x_{1}Q^{\alpha
}_{\perp}(k_1-x_{1}k')^{\beta}_{\perp}
+x_{2}(k_1-x_{1}k')^{\alpha}_{\perp}Q^{\beta}_{\perp}
}{(k_1-x_{1}k')^{2}_{\perp}}-
\frac{x_{1}Q^{\alpha}_{\perp}(k_2-x_{2}k')^{\beta}_{\perp}
+x_{2}(k_2-x_{2}k')^{\alpha}_{\perp}Q^{\beta}_{\perp}}
{(k_2-x_{2}k')^{2}_{\perp}}~,\ee
\be {\cal L}^{\alpha\beta}_{b1}=-b^{\alpha
\beta}(q_{1};k'_{1},k_{2}), \ee
\be{\cal L}^{\alpha\beta}_{c1}
=-b^{\beta\alpha}(q_{1};k'_{2},k_{1}), \ee
\be {\cal
L}^{\alpha\beta}_{d1}=\frac{q^{2}_{1\perp}}{2k'^{2}_{\perp}} \Bigg[
b^{\alpha\beta}(q_{1}-r;k_{1},k_{2})+{b}^{\beta\alpha}(q_{1}-r;k_{2},k_{1})\Bigg],
\ee
\be {\cal L}^{\alpha\beta}_{f1}=
-\frac{q^{2}_{1\perp}}{2(q_1-k'_{1})^{2}_{\perp}k'^2_{1\perp}}
(k'_{1}-k_{1}\frac{{k'}^{2}_{1\perp}}{k^{2}_{1\perp}})^{\alpha}
_{\perp}(q_1-k'_{1}-k_{2}\frac{(q_1-k'_{1})^2_{\perp}}
{k^2_{2\perp}})^{\beta}_{\perp}. \ee Here and below
$Q_\perp=(q_1-k'{q^{2}_{1\perp}}/{k'^2_\perp})_\perp $. According
to (\ref{boot L}) the integrated sum of ${\cal
L}^{\alpha\beta}_{m1}$ must be zero. One can track the
cancellation of separate contributions  using the decompositions
(\ref{eq:denom1}),(\ref{eq:denom2}) and the change of variables $
r_\perp \leftrightarrow (q_2-r)_\perp $, at which
 $D(q_1-k'_1,k_1)\leftrightarrow D(k'_2,q_1-k'_2)$  and consequently,
\begin{equation}
\frac{-x_{1}q^{\alpha}_{1\perp}
(q_1-k'_1)^{\beta}_{\perp}+x_2k'^{\alpha}_{1\perp}q^{\beta}_{1}}{D(q_1-k'_1,k'_1)}
\leftrightarrow \frac{-x_1
k'^{\beta}_{2\perp}q^{\alpha}_{1\perp}+x_{2}q^{\beta}_{1\perp}(q_1-k'_2)^{\alpha}_{\perp}
} {  D(k'_2,q_1-k'_2)  }~.
\end{equation}
After that  the cancellation of the terms with
$g^{\alpha\beta}_{\perp}$ follows from  trivial relations:
\begin{equation}
2x_1k'_{\perp}(k_1-x_1k')_{\perp}-k^2_{1\perp}=-(d(k'_2,k_1)+x^2_1k'^2_{\perp}),
\label{help3}
\end{equation}
\begin{equation}
2x_2k'_{\perp}(k_2-x_2k')_{\perp}-k^2_{2\perp}=-(d(k_2,k'_1)+x^2_2k'^2_{\perp}).
\label{help4}
\end{equation}
To see that the sum  of all other  terms is zero the equality
\begin{equation}
\int\frac{d^{D-2}r_\perp}{r_\perp^2(q_2-r)^{2}_\perp}[x_2(1-
\frac{k'^2_{1\perp}}{D(q_1-k'_1,k'_1)})+x_1(1-\frac{k'^{2}_2}{D(k'_2,q_1-k'_2)})]=0~,
\label{help5}
\end{equation}
which follows from the change of variables $ r_\perp
\leftrightarrow (q_2-r)_\perp $  and $x_1+x_2=1$, is helpful.

Let us turn to the case $n=2$ in  (\ref{boot L}). For separate
terms in the integrand we obtain
\[ {\cal L}^{\alpha\beta}_{a2}=
-x_1x_2g^{\alpha\beta}_{\perp}\Bigg[\frac{Q_{\perp}(k_2-x_2k')_{\perp}}{d(k_2,k'_1)}+
\frac{Q_{\perp}(x_2k_1-x_1k_2)_{\perp}}{d(k_2,k_1)} \Bigg]
\]
\be +\frac{x_{1}Q^{\alpha
}_{\perp}(x_2k_1-x_{1}k_2)^{\beta}_{\perp}+x_{2}(x_2k_1-x_{1}k_2)^{\alpha}_{\perp}Q^{\beta}_{\perp}
}{d(k_2,k_1)}+\frac{x_{1}Q^{\alpha
}_{\perp}(k_2-x_{2}k')^{\beta}_{\perp}
+x_{2}(k_2-x_{2}k')^{\alpha}_{\perp}Q^{\beta}_{\perp}
}{d(k_2,k'_1)}~,\ee
\be {\cal L}^{\alpha\beta}_{b2}= b^{\alpha\beta}(q_1;k'_1,k_2), \ee
\be {\cal L}^{\alpha\beta}_{c2}={b}^{\beta\alpha}(q_1;k'_2,k_1)
+b^{\alpha\beta}(q_1;k_1,k'_2), \ee
\be {\cal
L}^{\alpha\beta}_{d2}=-\frac{q^{2}_{1\perp}}{2k'^2_{\perp}}{b}^{\beta\alpha}
(q_1-r;k_2,k_1), \ee
\be {\cal L}^{\alpha\beta}_{f2}=
\frac{q^2_{1\perp}}{2(q_1-k'_1)^2_{\perp}k'^2_{1\perp}}(k'_1-k_1\frac{k'^2_{1\perp}}
{k^2_{1\perp}})^{\alpha}_{\perp}(q_1-k'_1-k_2\frac{(q_1-k'_1)^2_{\perp}}
{k^2_{2\perp}})^{\beta}_{\perp}~. \ee Although separate
contributions  in (\ref{boot L}) are rather complicated,  their
sum can be greatly simplified using the equalities
\begin{equation}
\int
\frac{d^{D-2}r_\perp}{r_{\perp}^2(q_2-r)^{2}_\perp}\Bigg[\frac{x_1k'^{\alpha}_{1\perp}(q_1-k'_1)^{\beta}_{\perp}}{k'^2_{1\perp}D(q_1-k'_1,k'_1)}+\frac{x_2k'^{\beta}_{2\perp}(q_1-k'_2)^{\alpha}_{\perp}}{k'^2_{2\perp}D(k'_2,q_1-k'_2)}-\frac{k'^{\alpha}_{1\perp}(q_1-k'_1)^{\beta}_{\perp}}{k'^2_{1\perp}(q_1-k'_1)^2_{\perp}}\Bigg]
= 0,
\end{equation}
\begin{equation}
\int\frac{d^{D-2}r_\perp}{r_{\perp}^2(q_2-r)^{2}_\perp}\Bigg[\frac{x_1k^{\alpha}_{1\perp}k'^{\beta}_{2\perp}}{k^2_{1\perp}D(k'_2,k_1)}+\frac{x_2k^{\alpha}_{1\perp}k'^{\beta}_{2\perp}}{k'^2_{2\perp}D(k'_2,k_1)}-\frac{k^{\alpha}_{1\perp}(q_1-k'_1)^{\beta}_{\perp}}{k^{2}_{1\perp}(q_1-k'_1)^{2}_{\perp}}\Bigg]
= 0,
\end{equation}
which are readily follow from the change of variables $ r_\perp
\leftrightarrow (q_2-r)_\perp $, relations
(\ref{help3}),(\ref{help4}) and not less trivial equality
\begin{equation}
x_2k'^{\beta}_{\perp}k'^{\alpha}_{1\perp}-x_1k^{\beta}_{2\perp}k'^{\alpha}_{\perp}=k'^{\alpha}_{1\perp}(x_2k'_1-x_1k_2)^{\beta}_{\perp}+k^{\beta}_{2\perp}(x_2k'_1-x_1k_2)^{\alpha}_{\perp}.
\label{help6}
\end{equation}
After that fulfillment of (\ref{boot L}) for $n=2$ becomes plain.

Finally, consider (\ref{boot L}) at $n=3$.   For the coefficients
${\cal L}^{\alpha\beta}_{m3}$ we obtain:
\par
\[{\cal
L}^{\alpha\beta}_{a3}=x_1x_2g^{\alpha\beta}_{\perp}\Bigg[\frac{Q_{\perp}(k_2-x_2k')_{\perp}}{d(k_2,k'_1)}
+\frac{Q_{\perp}(x_2k_1-x_1k_2)_{\perp}}{d(k_2,k_1)} \Bigg]
\]
\be -\frac{x_1 Q^{\alpha}_{\perp}(k_2-x_2k')^{\beta}_{\perp}+x_2
Q^{\beta}_{\perp}(k_2-x_2k')^{\alpha}_{\perp}}{d(k_2,k'_1)}-\frac{x_1
Q^{\alpha}_{\perp}(x_2k_1-x_1k_2)^{\beta}_{\perp}+x_2
Q^{\beta}_{\perp}(x_2k_1-x_1k_2)^{\alpha}_{\perp}}{d(k_2,k_1)}~,
\ee
\be {\cal L}^{\alpha\beta}_{b3}={b}^{\beta\alpha}(q_1;k_2,k'_1),
\ee
\be {\cal L}^{\alpha\beta}_{c3}=0, \ee
\be {\cal
L}^{\alpha\beta}_{d3}=\frac{q^2_{1\perp}}{2k'^2_{\perp}}{b}^{\beta\alpha}(q_1-r;k_2,k_1),
\ee
\be {\cal L}^{\alpha\beta}_{f3}=0, \ee  Verification of (\ref{boot
L}) is rather simple here; the trivial equality
\begin{equation}
2x_1k'_{\perp}(k_2-x_2k')_{\perp}+k'^2_{1\perp}=d(k_2,k'_1)+x^2_1k'^2_{\perp}.
\end{equation}
is helpful to perform it.
\section{Summary and discussion} In this paper we have calculated
in the one-loop approximation the leading logarithmic corrections
to the QCD amplitudes in the QMRK.   We have considered two
essentially different kinematics. In one of them two particles
with limited invariant mass are produced in the fragmentation
region of one of colliding particles. In another there are two
gaps between rapidities of the produced particles and rapidities
of colliding ones (production in the central region).  The
radiative corrections were calculated using the $s$-channel
unitarity.  In both cases we have found that the radiative
corrections are just the same which are prescribed by the
Reggeized form of the amplitudes. It is worth-while to note that
this form of corrections appears as a result of miraculous
cancellations between various contributions. The $s$-channel
unitarity method used by us for the calculation is very economic.
Using this method we have to consider only a few contributions,
whereas number of Feynman diagrams is estimated by hundreds.
Nevertheless, even in this approach the cancellations are quite
impressive.

Since in the $s$-channel unitarity method the radiative
corrections are expressed in terms of the Reggeon vertices, the
cancellation appears as a result of fulfillment of Eqs. (\ref{boot
qmrk}) and (\ref{boot central}). Therefore these equations are the
bootstrap conditions, necessary for compatibility of the Reggeized
form of the amplitudes with the $s$-channel unitarity.

The gluon Reggeization is one of remarkable properties of QCD,
very important at high energies. It is proved in the LLA, but
still remains a hypothesis in the NLA. This hypothesis can be
checked, and, hopefully, proved~\cite{tbp}  using the bootstrap
requirement, i.e. the demand of compatibility of the Reggeized
form of the amplitudes with the $s$-channel unitarity. The
requirement leads to an infinite set of the bootstrap relations
for the scattering amplitudes.  Fulfillment of these relations
guarantees the Reggeized form of the radiative corrections order
by order in perturbation theory. It occurs that all these
relations can be satisfied if the Reggeon vertices and the gluon
Regge trajectory submit to several bootstrap conditions. The proof
of the gluon Reggeization in the LLA~\cite{BLF} is just
demonstration that fulfilment of the bootstrap conditions in the
leading order is sufficient to satisfy all bootstrap relations.
Hopefully, the same can be done in the NLA~\cite{tbp}. There are
no doubts that the Reggeized form of the QMRK amplitudes can be
proved in such way. Since these amplitudes contain the gluon Regge
trajectory and the Reggeon-Reggeon-gluon vertex in the leading
order, the only new (compared with the LLA) thing which is
required to perform the proof is fulfillment of the bootstrap
conditions (\ref{boot qmrk}) and (\ref{boot central}). We'll
return to this question elsewhere.

\vspace{0.2cm} \noindent \underline{\bf Acknowledgments:} One of
us (V.S.F.)  thanks the Alexander von Humboldt foundation for the
research award, the Universit\"at Hamburg and DESY for their warm
hospitality while a part of this work was done.

\end{document}